\documentclass[11pt,a4paper]{article}
\pdfoutput=1
\usepackage{jheppub}
\usepackage[T1]{fontenc}
\usepackage{float}
\usepackage{color}
\usepackage{epic}
\begin{document}
\preprint{IP/BBSR/2015-4}
\title{Neutrinoless Double Beta Decay in Type I+II Seesaw Models}
\author[a,1]{Debasish Borah \note{Corresponding author}}
\affiliation[a]{Department of Physics, Tezpur University, Tezpur-784028, India}
\author[b]{Arnab Dasgupta}
\affiliation[b]{Institute of Physics, Sachivalaya Marg, Bhubaneshwar-751005, India}
%\emailAdd{dborah@tezu.ernet.in}
%\emailAdd{mkdas@tezu.ernet.in}

\emailAdd{dborah@tezu.ernet.in}
\emailAdd{arnab.d@iopb.res.in}

\abstract{We study neutrinoless double beta decay in left-right symmetric extension of the standard model with type I and type II seesaw origin of neutrino masses. Due to the enhanced gauge symmetry as well as extended scalar sector, there are several new physics sources of neutrinoless double beta decay in this model. Ignoring the left-right gauge boson mixing and heavy-light neutrino mixing, we first compute the contributions to neutrinoless double beta decay for type I and type II dominant seesaw separately and compare with the standard light neutrino contributions. We then repeat the exercise by considering the presence of both type I and type II seesaw, having non-negligible contributions to light neutrino masses and show the difference in results from individual seesaw cases. Assuming the new gauge bosons and scalars to be around a TeV, we constrain different parameters of the model including both heavy and light neutrino masses from the requirement of keeping the new physics contribution to neutrinoless double beta decay amplitude below the upper limit set by the GERDA experiment and also satisfying bounds from lepton flavor violation, cosmology and colliders.}

\maketitle

\section{Introduction}
The Standard Model (SM) of particle physics is still the most successful theory of elementary particles and their interactions except gravity. After the discovery of its last missing piece, the Higgs boson at the large hadron collider (LHC) in 2012, no convincing sign of new physics has appeared until the 8 TeV run of LHC. Though these null results are adding more feathers to the SM cap, the particle physics community is becoming desperate to discover some new physics beyond the SM (BSM). This is due to the severe inadequacies in the SM as it fails to explain many experimentally observed phenomena and address some theoretical questions. One such observed phenomena is the tiny but non-zero neutrino masses and large neutrino mixing \cite{PDG}. Due to the absence of right handed neutrinos, one can not write down a Dirac mass term for the neutrinos whereas the Majorana mass term for the neutrinos are disallowed by the gauge structure of SM. This keeps the neutrinos massless in SM with zero mixing among them which is ruled out by the recent neutrino experiments T2K \cite{T2K}, Double ChooZ \cite{chooz}, Daya-Bay \cite{daya} and RENO \cite{reno} . These recent experiments have not only made the earlier measurements of neutrino parameters more precise but also led to the discovery of non-zero reactor mixing angle $\theta_{13}$ which was considered to be (very close) zero earlier. The $3\sigma$ global fit values of neutrino oscillation parameters that have appeared in the recent analysis of \cite{schwetz14}  and \cite{valle14}  are shown in table \ref{tab:data1} and \ref{tab:data2} respectively.
\begin{center}
\begin{table}[htb]
\begin{tabular}{|c|c|c|}
\hline
Parameters & Normal Hierarchy (NH) & Inverted Hierarchy (IH) \\
\hline
$ \frac{\Delta m_{21}^2}{10^{-5} \text{eV}^2}$ & $7.02-8.09$ & $7.02-8.09 $ \\
$ \frac{|\Delta m_{31}^2|}{10^{-3} \text{eV}^2}$ & $2.317-2.607$ & $2.307-2.590 $ \\
$ \sin^2\theta_{12} $ &  $0.270-0.344 $ & $0.270-0.344 $ \\
$ \sin^2\theta_{23} $ & $0.382-0.643$ &  $0.389-0.644 $ \\
$\sin^2\theta_{13} $ & $0.0186-0.0250$ & $0.0188-0.0251 $ \\
$ \delta $ & $0-2\pi$ & $0-2\pi$ \\
\hline
\end{tabular}
\caption{Global fit $3\sigma$ values of neutrino oscillation parameters \cite{schwetz14}}
\label{tab:data1}
\end{table}
\end{center}
\begin{center}
\begin{table}[htb]
\begin{tabular}{|c|c|c|}
\hline
Parameters & Normal Hierarchy (NH) & Inverted Hierarchy (IH) \\
\hline
$ \frac{\Delta m_{21}^2}{10^{-5} \text{eV}^2}$ & $7.11-8.18$ & $7.11-8.18 $ \\
$ \frac{|\Delta m_{31}^2|}{10^{-3} \text{eV}^2}$ & $2.30-2.65$ & $2.20-2.54 $ \\
$ \sin^2\theta_{12} $ &  $0.278-0.375 $ & $0.278-0.375 $ \\
$ \sin^2\theta_{23} $ & $0.393-0.643$ &  $0.403-0.640 $ \\
$\sin^2\theta_{13} $ & $0.0190-0.0262$ & $0.0193-0.0265 $ \\
$ \delta $ & $0-2\pi$ & $0-2\pi$ \\
\hline
\end{tabular}
\caption{Global fit $3\sigma$ values of neutrino oscillation parameters \cite{valle14}}
\label{tab:data2}
\end{table}
\end{center}
Although the $3\sigma$ range for the leptonic Dirac CP phase $\delta$ is $0-2\pi$, there are two possible best fit values of it found in the literature: $306^o$ (NH), $254^o$ (IH) \cite{schwetz14} and $254^o$ (NH), $266^o$ (IH) \cite{valle14}. There has also been a hint of this Dirac phase to be $-\pi/2$ as reported by \cite{diracphase} recently. Although the absolute mass scale of the neutrinos are not yet known, we have an upper bound on the sum of absolute neutrino masses from cosmology, given by the Planck experiment $\sum_i \lvert m_i \rvert < 0.23$ eV \cite{Planck13}.

This observation of non-zero but tiny neutrino masses and mixing have led to a significant number of research activities in the last few decades in the form of several well motivated BSM frameworks. The fact that the neutrino masses are found to lie at least twelve order of magnitude lower than the electroweak scale, and the pattern of neutrino mixing with large mixing angles is very different from quark mixing with small mixing angles, gives the hint that their origin must be different from the physics at electroweak scale. The most popular BSM framework explaining the tiny sub-eV neutrino masses is the seesaw mechanism which broadly fall into three categories namely, type I \cite{ti}, type II \cite{tii0,tii} and type III \cite{tiii}, all of which involve the introduction of additional heavy fermion or scalar particles into the SM. In generic seesaw models, there exists a hierarchy between the electroweak scale and the scale of heavy fermions or scalars required to arrive at the suppression for neutrino masses. Therefore in typical seesaw models with order one dimensionless couplings, the additional massive fermion or scalar fields lie at a scale much beyond the energies accessible to present experiments like LHC. Bringing these additional particles to the TeV ballpark involves the fine-tuning of Yukawa couplings so as to keep the neutrino masses at sub-eV scale. In TeV scale type I seesaw, the Dirac Yukawa couplings have to be fine tuned to at least $10^{-6}-10^{-5}$, thereby reducing their production cross sections at colliders. TeV scale type II and type III seesaw have slightly better scope of having collider signatures due to the presence of electroweak gauge interactions of the additional particles. Although the ongoing LHC experiment is still hunting for such new physics signatures, it is equally important to look for some independent probe of these seesaw models. One such promising arena is the neutrinoless double beta decay (NDBD). For a review, please see \cite{NDBDrev}.

Neutrinoless double beta decay is a process where a nucleus emits two electrons thereby changing its atomic number by two units
$$ (A, Z) \rightarrow (A, Z+2) + 2e^- $$
with no neutrinos in the final state. Such a process violates lepton number by two units and hence is a probe of Majorana neutrinos, which are predicted by generic seesaw models of neutrino masses. There have been serious experimental efforts to detect such a process in the last few years. The latest experiments that have improved the lower bound on the half-life of NDBD process are KamLAND-Zen \cite{kamland} and GERDA \cite{GERDA} using Xenon-136 and Germanium-76 nuclei respectively. The light SM neutrino contribution to the half-life of NDBD can be as small as the lower bound set by these experiments only for quasi-degenerate type light neutrino spectrum. Such a mass spectrum of light neutrinos is however, tightly constrained from the Planck upper bound on the sum of absolute neutrino masses mentioned above. A future observation of NDBD with a half-life close to the present lower bound could therefore be a sign of new physics as the Planck upper bound on $\sum_i \lvert m_i \rvert$ may not allow light SM neutrino explanation for the same.  Here we consider a combination of type I and type II seesaw mechanisms as the new physics part which, apart from giving rise to light neutrino masses, can also give additional contributions to NDBD half-life. We also extend the SM gauge symmetry to that of Left-Right Symmetric Models (LRSM) \cite{lrsm} where such type I+II seesaw arises naturally. Several earlier works \cite{tii0, NDBDprev} have calculated the new physics contributions to NDBD within this model. More recently, the authors of \cite{ndbd00, ndbd0} studied the new physics contributions to NDBD process for TeV scale LRSM with dominant type II seesaw. There have also been several works \cite{ndbd1} where type I seesaw limit was also included into the computation of NDBD in LRSM. Some more detailed analysis incorporating left-right gauge mixing were discussed in the works \cite{ndbd2}, both in minimal as well as non-minimal versions of LRSM.

In the present work, we consider a scenario where both type I and type II seesaw terms can be equally dominating which to our knowledge, was not discussed previously in the context of NDBD. Such a combination of type I and type II seesaw together can also explain non-zero reactor mixing angle $\theta_{13}$ as well as CP phases, as discussed in the works \cite{db-t2}. Making use of the presence of two equally dominating seesaw terms in the neutrino mass formula, we consider the possibility where either type I or type II or both type I and type II mass matrices can be arbitrary while structural cancellation \cite{structure} between them can give rise to the correct light neutrino mass matrix. Instead of considering completely arbitrary type I and type II mass matrices, we consider a specific type I seesaw mass matrix: tri-bimaximal (TBM) type which gives $\theta_{13}=0$ \cite{db-t2}. The type II seesaw mass matrix is then constructed in such a way that the correct light neutrino mass matrix is obtained. Assuming the left-right symmetry breaking scale to be within the TeV range, we then constrain the parameters of the model as well as the relative contribution of individual seesaw terms to light neutrino mass formula from the requirement of keeping the NDBD amplitude below the upper bound set by experiments. We find that the new physics contribution can be quite close to or even above this upper bound for some region of parameter space even if the light neutrino spectrum is not quasi-degenerate type. This not only allow us to rule out some region of parameter space but also increases the possibility of detecting such a process in successive run of double beta decay experiments like GERDA. 

This paper is organized as follows. In section \ref{sec1}, we briefly discuss the left-right symmetric model and then summarize the origin of neutrino masses in this model in section \ref{sec2}. In section \ref{sec1a}, we briefly point out the possible new physics sources to neutrinoless double beta decay amplitude and discuss them in the limit of type I seesaw dominance, type II seesaw dominance and equal dominance of both type I and type II seesaw. In section \ref{sec:lfvlhc} we briefly discuss the existing experimental bounds on the masses of new particles in the model. In section \ref{sec3}, we discuss our numerical analysis and finally conclude in \ref{sec4}.

\section{Left-Right Symmetric Model}
\label{sec1}
Left-Right Symmetric Model \cite{lrsm} is one of the most popular BSM framework where the gauge symmetry of the SM is extended to $SU(3)_c \times SU(2)_L \times SU(2)_R \times U(1)_{B-L}$. The right handed fermions which are singlets under the $SU(2)_L$ of SM, transform as doublets under $SU(2)_R$, making the presence of right handed neutrinos natural in this model. The Higgs doublet of the SM is replaced by a Higgs bidoublet to allow couplings between left and right handed fermions, both of which are doublets under $SU(2)_L$ and $SU(2)_R$ respectively. The enhanced gauge symmetry of the model $SU(2)_R \times U(1)_{B-L}$ is broken down to the $U(1)_Y$ of SM by the vacuum expectation value (vev) of additional Higgs scalar, transforming as triplet under $SU(2)_R$ and having non-zero $U(1)_{B-L}$ charge. This triplet also gives Majorana masses to right handed neutrinos, responsible for type I seesaw. The left handed Higgs triplet on the other hand, can give tiny Majorana masses to the SM neutrinos through type II seesaw mechanism.

The fermion content of the minimal LRSM is
\begin{equation}
Q_L=
\left(\begin{array}{c}
\ u_L \\
\ d_L
\end{array}\right)
\sim (3,2,1,\frac{1}{3}),\hspace*{0.8cm}
Q_R=
\left(\begin{array}{c}
\ u_R \\
\ d_R
\end{array}\right)
\sim (3^*,1,2,\frac{1}{3}),\nonumber 
\end{equation}
\begin{equation}
\ell_L =
\left(\begin{array}{c}
\ \nu_L \\
\ e_L
\end{array}\right)
\sim (1,2,1,-1), \quad
\ell_R=
\left(\begin{array}{c}
\ \nu_R \\
\ e_R
\end{array}\right)
\sim (1,1,2,-1) \nonumber
\end{equation}
Similarly, the Higgs content of the minimal LRSM is
\begin{equation}
\Phi=
\left(\begin{array}{cc}
\ \phi^0_{11} & \phi^+_{11} \\
\ \phi^-_{12} & \phi^0_{12}
\end{array}\right)
\sim (1,2,2,0)
\nonumber 
\end{equation}
\begin{equation}
\Delta_L =
\left(\begin{array}{cc}
\ \delta^+_L/\surd 2 & \delta^{++}_L \\
\ \delta^0_L & -\delta^+_L/\surd 2
\end{array}\right)
\sim (1,3,1,2), \hspace*{0.2cm}
\Delta_R =
\left(\begin{array}{cc}
\ \delta^+_R/\surd 2 & \delta^{++}_R \\
\ \delta^0_R & -\delta^+_R/\surd 2
\end{array}\right)
\sim (1,1,3,2) \nonumber
\end{equation}
where the numbers in brackets correspond to the quantum numbers with respect to the gauge group $SU(3)_c\times SU(2)_L\times SU(2)_R \times U(1)_{B-L}$. In the symmetry breaking
pattern, the neutral component of the Higgs triplet $\Delta_R$ acquires a vev to break the gauge symmetry of the LRSM into that of the SM and then to the $U(1)$ of electromagnetism by the vev of the neutral component of Higgs bidoublet $\Phi$:
$$ SU(2)_L \times SU(2)_R \times U(1)_{B-L} \quad \underrightarrow{\langle
\Delta_R \rangle} \quad SU(2)_L\times U(1)_Y \quad \underrightarrow{\langle \Phi \rangle} \quad U(1)_{em}$$
The symmetry breaking of $SU(2)_R \times U(1)_{B-L}$ into the $U(1)_Y$ of standard model can also be achieved at two stages by choosing a non-minimal scalar sector \cite{lrdb}.

\section{Neutrino Mass in LRSM}
\label{sec2}
The relevant Yukawa couplings which lead to small non-zero neutrino
mass are given by
\begin{eqnarray}
{\cal L}^{II}_\nu &=& y_{ij} \ell_{iL} \Phi \ell_{jR}+ y^\prime_{ij} \ell_{iL}
\tilde{\Phi} \ell_{jR} +h.c.
\nonumber \\
&+& f_{ij}\ \left(\ell_{iR}^T \ C \ i \sigma_2 \Delta_R \ell_{jR}+
(R \leftrightarrow L)\right)+h.c.
\label{treeY}
\end{eqnarray}
where $\tilde{\Phi} = \tau_2 \Phi^* \tau_2$. In the above Yukawa Lagrangian, the indices $i, j = 1, 2, 3$ correspond to the three generations of fermions. The Majorana Yukawa couplings $f$ is same for both left and right handed neutrinos
because of left-right symmetry $(f_L = f_R)$. These couplings $f$ give rise to the Majorana mass terms of both left handed and right handed neutrinos after the triplet Higgs fields $\Delta_{L,R}$ acquire non-zero vev. These mass terms appear in the seesaw formula as discussed below. The resulting seesaw formula in this minimal model can be written as
\begin{equation}
M_{\nu}=M_{\nu}^{II} + M_{\nu}^I
\label{type2a}
\end{equation}
where the usual type I seesaw term $M_{\nu}^I$ is given by the expression,
\begin{equation}
M_{\nu}^I=-m_{LR}M_{RR}^{-1}m_{LR}^{T}.
\end{equation}
Here  $m_{LR} = y v_1 + y^\prime v_2$ is the Dirac neutrino mass matrix, with $v_{1,2}$ are the vev's of the neutral components of the Higgs bidoublet. It should be noted that in the framework of LRSM, $M_{RR}$ arises naturally as a result of parity breaking at high energy and both the type I and type II terms can be written in terms of $M_{RR}$. In LRSM with Higgs triplets, $M_{RR}$ can be expressed as $M_{RR}=v_{R}f_{R}$. The first term $M_{\nu}^{II}$ in equation (\ref{type2a}) is due to the vev of $SU(2)_{L}$ Higgs triplet. Thus, $M_{\nu}^{II}=f_{L}v_{L}$ and $M_{RR}=f_{R}v_{R}$, where $v_{L,R}$ denote the vev's and $f_{L,R}$ are symmetric $3\times3$ matrices. The left-right symmetry demands $f_{R}=f_{L}=f$ as mentioned above. The induced vev for the left-handed triplet $v_{L}$ can be shown for generic LRSM to be
$$v_{L}=\gamma \frac{M^{2}_{W}}{v_{R}}$$
with $M_{W}\sim 80.4$ GeV being the weak boson mass such that 
$$ |v_{L}|<<M_{W}<<|v_{R}| $$ 
In general $\gamma$ is a function of various couplings in the scalar potential of generic LRSM. Using the results from Deshpande et al., (fifth reference in \cite{lrsm}), $\gamma$ is given by
\begin{equation}
\gamma = \frac{\beta_2 v^2_1+\beta_1 v_1 v_2 + \beta_3  v^2_2}{(2\rho_1-\rho_3)(v^2_1+v^2_2)}
\label{eq:gammaLR}
\end{equation}
where $\beta, \rho$ are dimensionless parameters of the scalar potential. Without any fine tuning $\gamma$ is expected to be of the order unity ($\gamma\sim 1$). However, for TeV scale type I+II seesaw, $\gamma$ has to be fine-tuned as we discuss later. The type II seesaw formula in equation (\ref{type2a}) can now be expressed as
\begin{equation}
M_{\nu}=\gamma (M_{W}/v_{R})^{2}M_{RR}-m_{LR}M^{-1}_{RR}m^{T}_{LR}
\label{type2}
\end{equation}

\begin{figure}[!h]
\centering
\begin{tabular}{ccc}
\epsfig{file=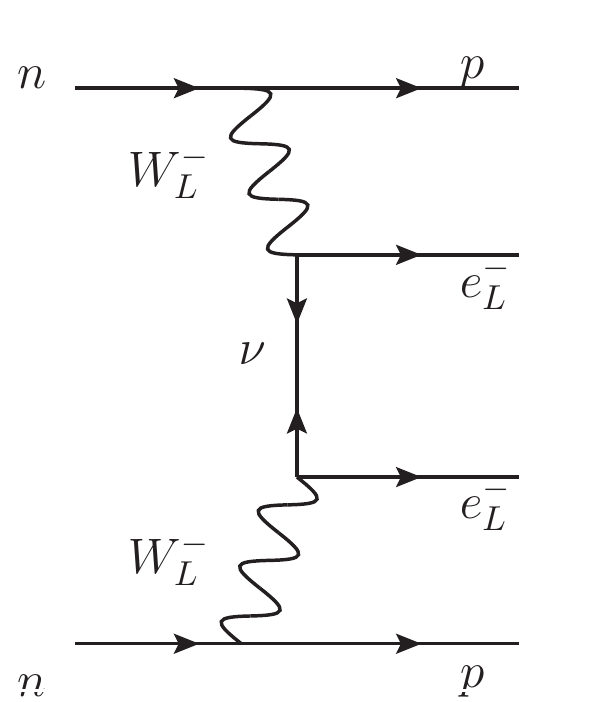,width=0.3\textwidth,clip=}&
\epsfig{file=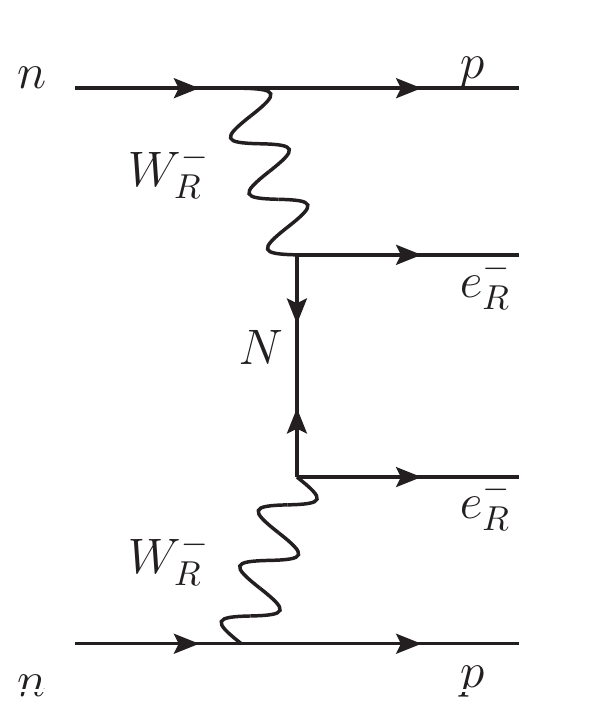,width=0.3\textwidth,clip=} &
\epsfig{file=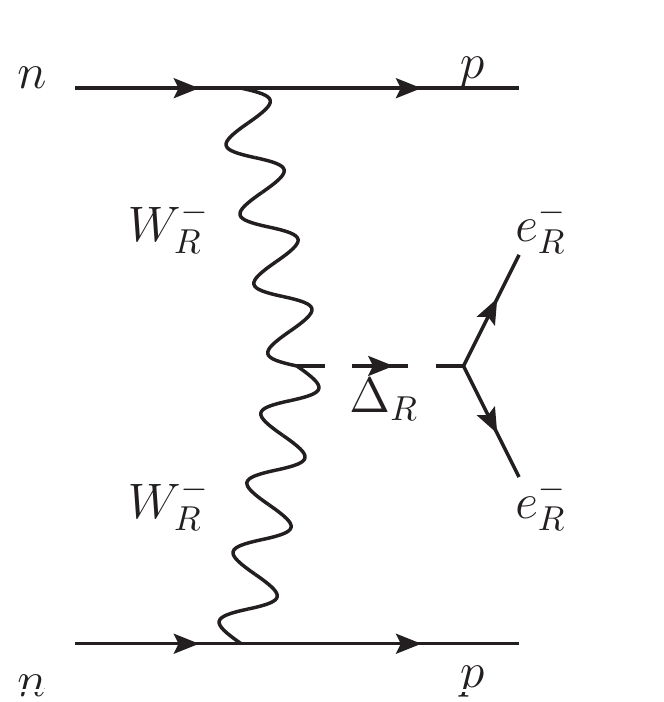,width=0.3\textwidth,clip=}
\end{tabular}
\caption{Feynman diagrams for Neutrinoless double beta decay due to $\nu-W_L, N-W_R, \Delta_R-W_R$ contributions}
\label{fig0}
\end{figure}

\section{NDBD in LRSM}
\label{sec1a}
Due to the presence of several additional vector, scalar and fermionic fields not present in the SM, one can have several new physics contributions to the neutrinoless double beta decay in LRSM. The corresponding Feynman diagrams are given in earlier works, for example \cite{ndbd1}. Here we consider only three contributions into account, as shown in figure \ref{fig0}.  Before discussing these three contributions in detail, we first briefly summarize the different possible contributions to NDBD in LRSM below.

1. The SM contribution comes from the Feynman diagram where the intermediate particles are $W^-_L$ bosons and light neutrinos. The amplitude of this process depends upon the leptonic mixing matrix elements and the light neutrino masses.

2. The light neutrino contribution can come also from the Feynman diagram mediated by $W^-_R$ bosons. The amplitude of this process depends upon the mixing between light and heavy neutrinos as well as $W^-_R$ mass. 

3. The light neutrino contribution can also come from the Feynman diagram mediated by both $W^-_L$ and $W^-_R$. The amplitude depends upon the mixing between light hand heavy neutrinos, leptonic mixing matrix elements, light neutrino masses and $W^-_R$ mass.

4. The heavy right handed neutrino $\nu_R$ contribution can come from the Feynman diagrams mediated by $W^-_L$ bosons. The amplitude depends upon the mixing between light and heavy neutrinos as well as masses of $\nu_R$.

5. The heavy right handed neutrino contribution can also come from the Feynman diagrams mediated by $W^-_R$ bosons. The corresponding amplitude depends upon the elements of right handed leptonic mixing matrix and masses of $\nu_R$. 

6. The heavy right handed neutrino contribution can come from the Feynman diagram where the intermediate particles are $W^-_L$ and $W^-_R$ simultaneously. The amplitude depends upon the right handed leptonic mixing elements, mixing between light and heavy neutrinos as well as heavy neutrino masses. 

7. The triplet Higgs scalars $\Delta_L$ and $\Delta_R$ can also contribute to neutrinos double beta decay through $W^-_L$ and $W^-_R$ mediation respectively. The amplitude depends upon the masses of $\Delta_{L,R}$ scalars as well as their couplings to leptons.

For the purpose of this work we consider only two contributions to NDBD in addition to the standard light neutrino contribution through $W^-_L$ exchange mentioned at point 1 above. These two corresponds to the ones mentioned at point 5 and point 7 above. We ignore the Feynman diagrams involving $W^-_L-W^-_R$ exchange as well as light and heavy neutrino mixings. We also ignore the mixing between $W^-_L$ and $W^-_R$ bosons, as it is tightly constrained from electroweak precision data. Among the triplet exchange diagrams, we only consider the one mediated by $W^-_R$ and $\Delta_R$ as the couplings. The other possible contribution through $\Delta_L, W^-_L$ is suppressed by the type II seesaw contribution to light neutrino masses and hence neglected here. The amplitude of the light neutrino contribution (first Feynman diagram in figure \ref{fig0}) considered here is 
\begin{equation}
A_\nu \propto \sum_i m_i U^2_{Lei} \mathcal{M}_\nu (0)
\end{equation}
where $\mathcal{M}_\nu (\mu)$ (following the notation of \cite{nme}) is the nuclear matrix element (NME) which is a function of the mediating neutrino mass $(\mu)$ for the NDBD of different nuclei. The NME above is written as $\mathcal{M}_\nu (0)$ as it is independent of neutrino mass for light neutrino exchange. In the above expression, $U_L$ is the leptonic mixing matrix which appears in left-handed charged current interactions and $m_i$ are the masses of light neutrinos for $i=1,2,3$.
%$$ A_\nu \propto \frac{U^2_{Lei}m_i}{p^2} $$
The contribution from the heavy neutrino and $W^-_R$ exchange (second Feynman diagram in figure \ref{fig0}) can be written as 
\begin{equation}
A_{RR} \propto \sum_i M_i U^{*2}_{Rei} \mathcal{M}^{RR}_\nu (M_i) = \left ( \frac{M_{W_L}}{M_{W_R}} \right )^4 \sum_i M_i U^{*2}_{Rei} \mathcal{M}_\nu (M_i)
\end{equation}
where $\mathcal{M}^{RR}_\nu = \left ( \frac{M_{W_L}}{M_{W_R}} \right )^4 \mathcal{M}_\nu$, $U_R$ is the right handed lepton mixing matrix and $M_i$ are the masses of right handed neutrinos for $i=1,2,3$.
%$$ A_{RR} \propto \frac{1}{M^4_{W_R}} \frac{U^2_{Rei}}{M_i} $$
The contribution from $W^-_R, \Delta_R$ exchange (third Feynman diagram in figure \ref{fig0}) is given by the amplitude
\begin{equation}
A_{R\Delta} \propto M^{ee}_{RR}\mathcal{M}^{\Delta}_\nu (M_{\Delta_R}) \approx \left ( \frac{M_{W_L}}{M_{W_R}} \right )^4 \frac{p^2}{M^2_{\Delta_R}} M^{ee}_{RR} \mathcal{M}_\nu (0)
\end{equation}
where $\frac{p^2}{M^2_{\Delta_R}}$ is the additional suppression coming from the scalar propagator with $p$ being the average momentum exchange for the process. In the above expression, $M^{ee}_{RR} =U^2_{Rei}M_i $ is the $(11)$ element of the right handed neutrino mass matrix.
%$$ A_{R\Delta} \propto \frac{1}{M^4_{W_R}} \frac{M^{ee}_{RR}}{M^2_{\Delta_R}} $$
Thus, the standard light neutrino contribution can be written as 
\begin{equation}
\frac{\Gamma^{\nu}_{\text{NDBD}}}{\text{ln}2} = G_F \frac{\lvert \mathcal{M}_\nu (0)\rvert^2}{m^2_e} \big \lvert U^2_{Lei}m_i \big \rvert^2 = G_F \frac{\lvert \mathcal{M}_\nu (0) \rvert^2}{m^2_e} \big \lvert m^{\text{eff}}_{\nu} \big \rvert^2
\label{eq:ndbd0}
\end{equation}
where $G_F$ is the Fermi coupling constant. Similarly, the new physics (NP) contributions considered in this model can be written as
\begin{equation}
\frac{\Gamma^{\text{NP}}_{\text{NDBD}}}{\text{ln}2} = G_F \frac{\lvert \mathcal{M}_\nu (0) \rvert^2}{m^2_e} \big \lvert \frac{M^4_{W_L}}{M^4_{W_R}} U^{*2}_{Rei} M_i \frac{\mathcal{M}_\nu (M_i)}{\mathcal{M}_\nu (0)}+p^2 \frac{M^4_{W_L}}{M^4_{W_R}} \frac{U^2_{Rei}M_i}{M^2_{\Delta_R}} \big \rvert^2
\nonumber
\end{equation}
which can further be simplified to get
\begin{equation}
\frac{\Gamma^{\text{NP}}_{\text{NDBD}}}{\text{ln}2} = G_F \frac{\lvert \mathcal{M}_\nu (0) \rvert^2}{m^2_e} \big \lvert p^2 \frac{M^4_{W_L}}{M^4_{W_R}} \frac{U^{*2}_{Rei}}{M_i}+p^2 \frac{M^4_{W_L}}{M^4_{W_R}} \frac{U^2_{Rei}M_i}{M^2_{\Delta_R}} \big \rvert^2
\label{eq:ndbd1}
\end{equation}
or in a shorter notation it can be written as
\begin{equation}
\frac{\Gamma^{\text{NP}}_{\text{NDBD}}}{\text{ln}2} = G \frac{\lvert \mathcal{M}_\nu (0)\rvert^2}{m^2_e} \big \lvert  m^{\text{eff}}_{N} + m^{\text{eff}}_{\Delta_R} \big \rvert^2
\label{eq:ndbd2}
\end{equation}
where $m^{\text{eff}}_{\nu}, m^{\text{eff}}_{N}, m^{\text{eff}}_{\Delta_R}$ are the effective neutrino masses corresponding to light neutrino $(\nu_L=\nu)$, heavy neutrino $(\nu_R=N)$ and triplet $(\Delta_R)$ contributions respectively to neutrinoless double beta decay. 

Our goal in this work is to point out the new physics contribution to NDBD when type I and type II seesaw both can be equally dominating. This can be very different from the type I or type II dominance cases discussed in earlier works, for example \cite{ndbd1}. To show this difference we have adopted the simplified approach of earlier work \cite{ndbd1} where the $W_L-W_R$ and $\nu-N$ mixing were neglected, resulting in the contributions to NDBD mentioned in equation \eqref{eq:ndbd1} being dominant. We leave a more general discussion including all possible contributions to a subsequent work.

Depending on the seesaw mechanism at work, these new physics sources can have different contributions to the neutrinoless double beta decay. We discuss type I dominance, type II dominance and equally dominant type I and type II seesaw mechanism below with reference to their contributions to neutrinoless double beta decay.

\subsection{Dominant Type I Seesaw}
For dominant type I seesaw, the first term on the right hand side of equation \eqref{type2} can be neglected. The light neutrino mass can then be written as
\begin{equation}
M_{\nu} = -\frac{m_{LR}f^{-1}_{R}m^{T}_{LR}}{v_R} = -\frac{1}{v_R} m_{LR} U_R f^{-1}_{\text{diag}} U^T_R m^T_{LR}
\end{equation}
where $U_R$ is the diagonalizing matrix of $M_{RR} = f_R v_R$. Multiplying both sides of $M_{\nu}$ by $U_R$ in the above equation, we get
$$ U^T_R M_{\nu} U_R = -\frac{1}{v_R} U^T_R m_{LR} U_R f^{-1}_{\text{diag}} U^T_R m^T_{LR} U_R$$
The most general Dirac neutrino mass matrix $m_{LR}$ is usually diagonalized by a bi-unitary transformation $m^{\text{diag}}_{LR} = U^T_L m_{LR} U_R$. Assuming $U_L=U_R$ gives 
$$ U^T_R M_{\nu} U_R = -\frac{1}{v_R} m^{\text{diag}}_{LR}f^{-1}_{\text{diag}}m^{\text{diag}}_{LR} = M^{\text{diag}}_{\nu} $$
Thus the light and heavy neutrino mass matrices diagonalized by the same unitary matrix $U_R$ in this approximation. Also the light neutrino masses are given by 
\begin{equation}
m_i = \frac{\left (m^{\text{diag}}_{LR} \right )^2_{ii}}{M_i}
\label{eq:mvsM}
\end{equation}
Thus, in the type I seesaw limit, we can replace the $U_R$ matrix in equation \eqref{eq:ndbd1} by the usual neutrino mixing matrix $U_L$. For diagonal charged lepton mass matrix, the neutrino mixing matrix $U_L$ is same as the leptonic mixing matrix, which is known from experimental data. The masses of $W^-_R, \Delta_R$ can be fixed within a few TeV, allowed by experimental constraints. One of the right handed neutrino masses $M_i$ can be fixed within a TeV whereas the others can be expressed as ratios of light neutrino masses due to the proportionality $m_i \propto \frac{1}{M_i}$ in the type I seesaw limit. For simplicity, the hierarchies of Dirac Yukawa couplings between different fermion generations can be neglected so that the heavy neutrino mass ratios can be written as mass ratios of light neutrinos, following from equation \eqref{eq:mvsM}.

\subsection{Dominant Type II Seesaw}
If the type I seesaw term is negligible, then the light neutrino is given by the first term on the right hand side of equation \eqref{type2}. Since the light neutrino mass matrix $M_{\nu}$ is proportional to the heavy neutrino mass matrix $M_{RR}$ in this case, the same unitary matrix can diagonalize both $M_{\nu}$ and $M_{RR}$. Also, in this case the light neutrino masses are directly proportional to heavy neutrino masses $m_i \propto M_i$ due to the proportionality between respective mass matrices. This allows us to express two heavy neutrino mass ratios in terms of light neutrino mass ratios. Thus, similar to the type I seesaw dominance case, here also we can write the new physics contribution to NDBD in terms of leptonic mixing matrix elements, one of the heavy neutrino masses, right handed gauge boson and right handed scalar triplet masses.

\subsection{Combination of Type I and Type II Seesaw}
The new physics contribution to NDBD can be different from the above two cases if type I and type II seesaw contributions to light neutrino masses are comparable. Some simple relations relating different mass matrices involved in the formula for light neutrino masses in LRSM given by equation \eqref{type2} were discussed in \cite{mdMRgoran}. One useful parametrization of the Dirac neutrino mass matrix in the presence of type I+II seesaw was studied by the authors of \cite{akhwer}. In another work \cite{akhfri}, relations between type I and type II seesaw mass matrices were derived by considering the Dirac neutrino mass matrix to be known. If the Dirac neutrino mass matrix $m_{LR}$ is not known, then we can still choose at least on of the type I and type II seesaw mass matrices arbitrarily due to the freedom we have in choosing $m_{LR}$ that appears in the type I seesaw term. After choosing one the seesaw mass matrices, the other gets completely fixed if the light neutrino mass matrix is completely known.

Instead of choosing arbitrary type I and type II seesaw mass matrices, it is really appealing for us to consider one of these seesaw mass matrices to possess $\mu-\tau$ symmetry or more specifically, Tri-Bimaximal or TBM type mixing. TBM mixing is a good approximation to observed neutrino mixing at leading order predicting the mixing angles as $\theta_{12} \simeq 35.3^o, \; \theta_{23} = 45^o$ and $\theta_{13} = 0$. This TBM mixing matrix discussed widely in the literature \cite{Harrison} can be written as
\begin{equation}
U_{\text{TBM}}=\left(\begin{array}{ccc}\sqrt{\frac{2}{3}}&\frac{1}{\sqrt{3}}&0\\
 -\frac{1}{\sqrt{6}}&\frac{1}{\sqrt{3}}&\frac{1}{\sqrt{2}}\\
\frac{1}{\sqrt{6}}&-\frac{1}{\sqrt{3}}& \frac{1}{\sqrt{2}}\end{array}\right)
\end{equation}
This TBM type mixing can be accommodated within several discrete flavor symmetry models \cite{discreteRev}. Since our intention in the present work is to do a phenomenological study of equally dominant type I and type II seesaw in the context of NDBD, we do not investigate the UV complete flavor symmetry framework of this scenario. The required correction to TBM type neutrino mass matrix in order to generate non-zero but small reactor mixing angle $\theta_{13}$ can originate from other seesaw terms or corrections from charged lepton sector. Here we consider a diagonal charged lepton mass matrix such that the leptonic mixing matrix is same as the diagonalizing matrix of the light neutrino mass matrix $M_{\nu}$. Although either type I or type II seesaw mass matrix can give rise to TBM type neutrino mixing, here we consider the type I seesaw mass matrix to be of TBM type mixing whereas type II seesaw mass matrix gives rise to the deviations from TBM in order to generate non-zero $\theta_{13}$. Since, type II seesaw term is proportional to the right handed neutrino mass matrix $M_{RR}$ in LRSM as shown in equation \eqref{type2}, one can construct $M_{RR}$ as a deviation from TBM form of type I seesaw mass matrix. One can perform this exercise the other way round as well that is, assuming the type II seesaw to give rise to TBM type neutrino mass matrix whereas type I seesaw gives the necessary correction. Although both of these approaches will produce the same light neutrino masses and mixing, they will have different implications in neutrinoless double beta decay. Here we consider only the former case, that is type I seesaw mass matrix of TBM type leaving the other possibility to future works.

The Pontecorvo-Maki-Nakagawa-Sakata (PMNS) leptonic mixing matrix is related to the diagonalizing 
matrices of neutrino and charged lepton mass matrices $U_{\nu}, U_l$ respectively, as
\begin{equation}
U_{\text{PMNS}} = U^{\dagger}_l U_{\nu}
\label{pmns0}
\end{equation}
The PMNS mixing matrix can be parametrized as
\begin{equation}
U_{\text{PMNS}}=\left(\begin{array}{ccc}
c_{12}c_{13}& s_{12}c_{13}& s_{13}e^{-i\delta}\\
-s_{12}c_{23}-c_{12}s_{23}s_{13}e^{i\delta}& c_{12}c_{23}-s_{12}s_{23}s_{13}e^{i\delta} & s_{23}c_{13} \\
s_{12}s_{23}-c_{12}c_{23}s_{13}e^{i\delta} & -c_{12}s_{23}-s_{12}c_{23}s_{13}e^{i\delta}& c_{23}c_{13}
\end{array}\right) U_{\text{Maj}}
\label{matrixPMNS}
\end{equation}
where $c_{ij} = \cos{\theta_{ij}}, \; s_{ij} = \sin{\theta_{ij}}$ and $\delta$ is the leptonic Dirac CP phase. The diagonal matrix $U_{\text{Maj}}=\text{diag}(1, e^{i\alpha}, e^{i(\beta+\delta)})$  contains the Majorana CP phases $\alpha, \beta$ which remain undetermined at neutrino oscillation experiments. In the diagonal $U_l$, the leptonic mixing matrix is $U_{\text{PMNS}} = U_{\nu} $ which for $U_{\nu} = U_{\text{TBM}}$ results in vanishing reactor mixing angle $\theta_{13}$ and the leptonic Dirac CP phase $\delta$. Thus, the type I seesaw mass matrix gives rise to vanishing $\theta_{13}$ and $\delta$ whereas type II seesaw mass matrix generates non-zero $\theta_{13}$ and non-trivial value of Dirac CP phase. Since the diagonalizing matrix of $M_{\nu}$ is $U_{\text{PMNS}}$ and that of type I mass matrix $M^I_{\nu}$ is $U_{\text{TBM}}$, the formula for light neutrino masses in the presence of type I and type II seesaw can be written as
\begin{equation}
U_{\text{PMNS}}M^{\text{diag}}_{\nu} U^T_{\text{PMNS}} = M^{II}_{\nu} -U_{\text{TBM}} U_{\text{Maj}}M^{I(\text{diag})}_{\nu} U^T_{\text{Maj}}U^T_{\text{TBM}}
\label{nu13}
\end{equation}
where the Majorana phases are incorporated in the type I seesaw term. In principle, the two terms on the right hand side of the above equation can have arbitrary strength provided the difference between them gives rise to the correct sub-eV scale light neutrino masses. The relative strength of type I and type II seesaw terms can be decided by introducing a parameter $X$ such that the diagonal type I seesaw mass matrix can be parametrized as $M^{I(\text{diag})}_{\nu} = X M^{\text{diag}}_{\nu}$. The parameter $X$ can take any numerical values, provided the two seesaw terms give rise to correct light neutrino mass matrix, after structural cancellation. For a particular value of $X$, one can construct the type II seesaw mass matrix using the above equation \eqref{nu13}. We denote the symmetric type II seesaw mass matrix as
\begin{equation}
M^{II}_{\nu}=\left(\begin{array}{ccc}
T_{11}& T_{12}&T_{13}\\
T_{12}& T_{22} & T_{23} \\
T_{13} & T_{23} & T_{33}
\end{array}\right)
\label{type2matrix}
\end{equation}
and using equation (\ref{nu13}), the type II seesaw mass matrix elements can be derived as shown in Appendix \ref{appendix1}.

For normal hierarchy, the diagonal mass matrix of the light neutrinos can be written  as $M^{\text{diag}}_{\nu} 
= \text{diag}(m_1, \sqrt{m^2_1+\Delta m_{21}^2}, \sqrt{m_1^2+\Delta m_{31}^2})$ whereas for inverted hierarchy 
 it can be written as $M^{\text{diag}}_{\nu} = \text{diag}(\sqrt{m_3^2+\Delta m_{23}^2-\Delta m_{21}^2}, 
\sqrt{m_3^2+\Delta m_{23}^2}, m_3)$. The mass squared differences can be taken from the global fit neutrino oscillation data shown in table \ref{tab:data1} and table \ref{tab:data2} shown above, leaving the lightest neutrino mass as free parameter in $M^{\text{diag}}$. Thus, the type II seesaw mass matrix can be written in terms of five free parameters: the lightest neutrino mass, three leptonic CP phases and the seesaw relative strength factor $X$. The right handed neutrino mass matrix can be found from 
$$ \gamma (M_{W}/v_{R})^{2}M_{RR} = M^{II}_{\nu}$$
which was also shown in equation \eqref{type2}. Thus, fixing the dimensionless parameter $X$ allows us to calculate the type II seesaw mass matrix in terms of neutrino parameters, which can then be used to find the right handed neutrino mass matrix $M_{RR}$ by fixing $\gamma$ and $v_R \approx M_{W_R}/g_R$.

\begin{table}[!h]
\centering
\begin{tabular}{|c|c|c|}
\hline
Parameters &  Values (NH) & Values (IH) \\
\hline
$\frac{\Delta m^2_{21}}{10^{-5} {\rm eV}^2}$ & 7.60  & 7.60\\
$\frac{\lvert \Delta m^2_{31}\rvert}{10^{-3} {\rm eV}^2} $ & 2.48 & 2.38 \\
%$\frac{\lvert \Delta m^2_{31}\rvert}{10^{-3} {\rm eV}^2}$ & 2.38 \\
$\sin^2 \theta_{12}$ & 0.323  & 0.323\\
$\sin^2 \theta_{23}$ & 0.567 & 0.573 \\
$\sin^2 \theta_{13}$ & 0.0234 & 0.024 \\
%$\delta$ & -$90^\circ$ & -$90^\circ$\\
%$M_{\Delta_R}$ & 1 TeV  & 1 TeV \\
%$M_{W_R}$ & 3 TeV & 3 TeV \\ 
$p$ & 100 MeV & 100 MeV \\ 
$M_{W_L}$ & 80.4 GeV & 80.4 GeV \\
\hline
\end{tabular}
\caption{Numerical values of several parameters used in the calculation of $m^{\text{eff}}$ for NDBD}
\label{param}
\end{table}
\begin{figure}[!h]
\centering
%\begin{tabular}{cc}
\epsfig{file=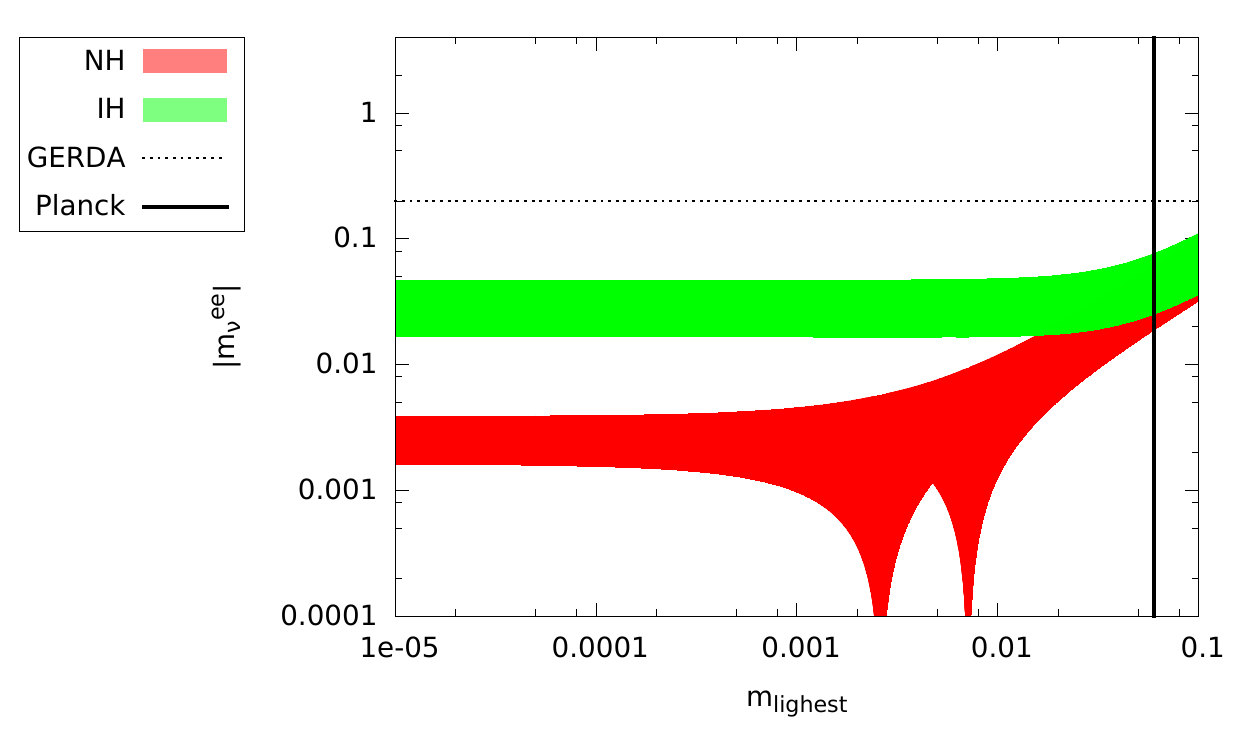,width=1.0\textwidth,clip=}
%\end{tabular}
\caption{Standard Model light neutrino contribution to effective neutrino mass which appears in NDBD}
\label{fig1}
\end{figure}
\begin{figure}[!h]
\centering
%\begin{tabular}{cc}
\epsfig{file=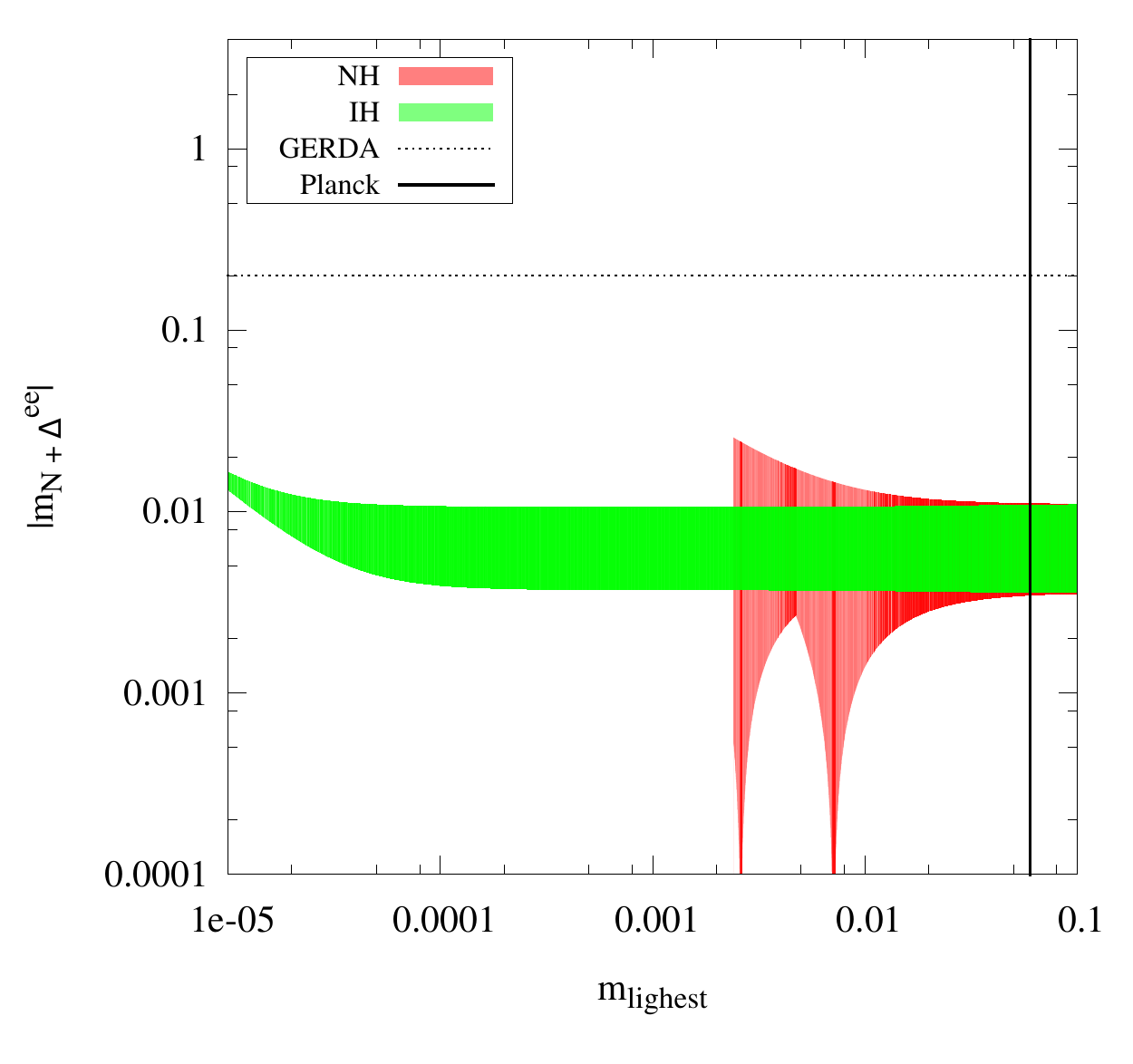,width=1.0\textwidth,clip=}
%\epsfig{file=TIMnD.pdf,width=0.5\textwidth,clip=} 
%\end{tabular}
\caption{New physics contribution to effective neutrino mass which appears in NDBD for the diagrams shown in figure \ref{fig0} with type I seesaw dominance and $M_{W_R}= M_{\Delta_R} = 3.5$ TeV.}
\label{fig2}
\end{figure}
\begin{figure}[!h]
\centering
%\begin{tabular}{cc}
\epsfig{file=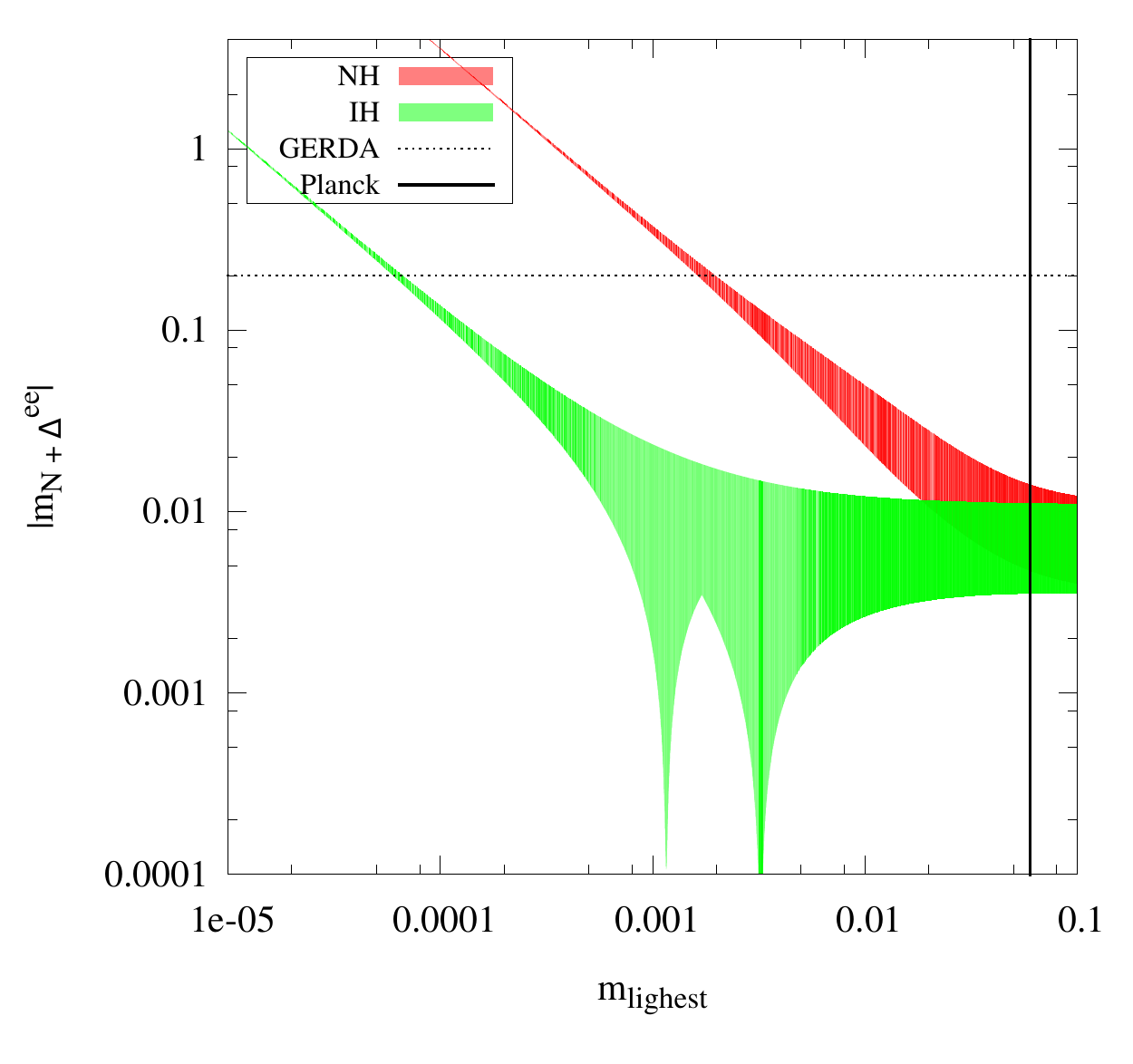,width=1.0\textwidth,clip=}
%\epsfig{file=TIIMnD.pdf,width=0.5\textwidth,clip=}
%\end{tabular}
\caption{New physics contribution to effective neutrino mass which appears in NDBD for the diagrams shown in figure \ref{fig0} with type II seesaw dominance and $M_{W_R}= M_{\Delta_R} = 3.5$ TeV.}
\label{fig3}
\end{figure}
\begin{figure}[!h]
\centering
%\begin{tabular}{cc}
\epsfig{file=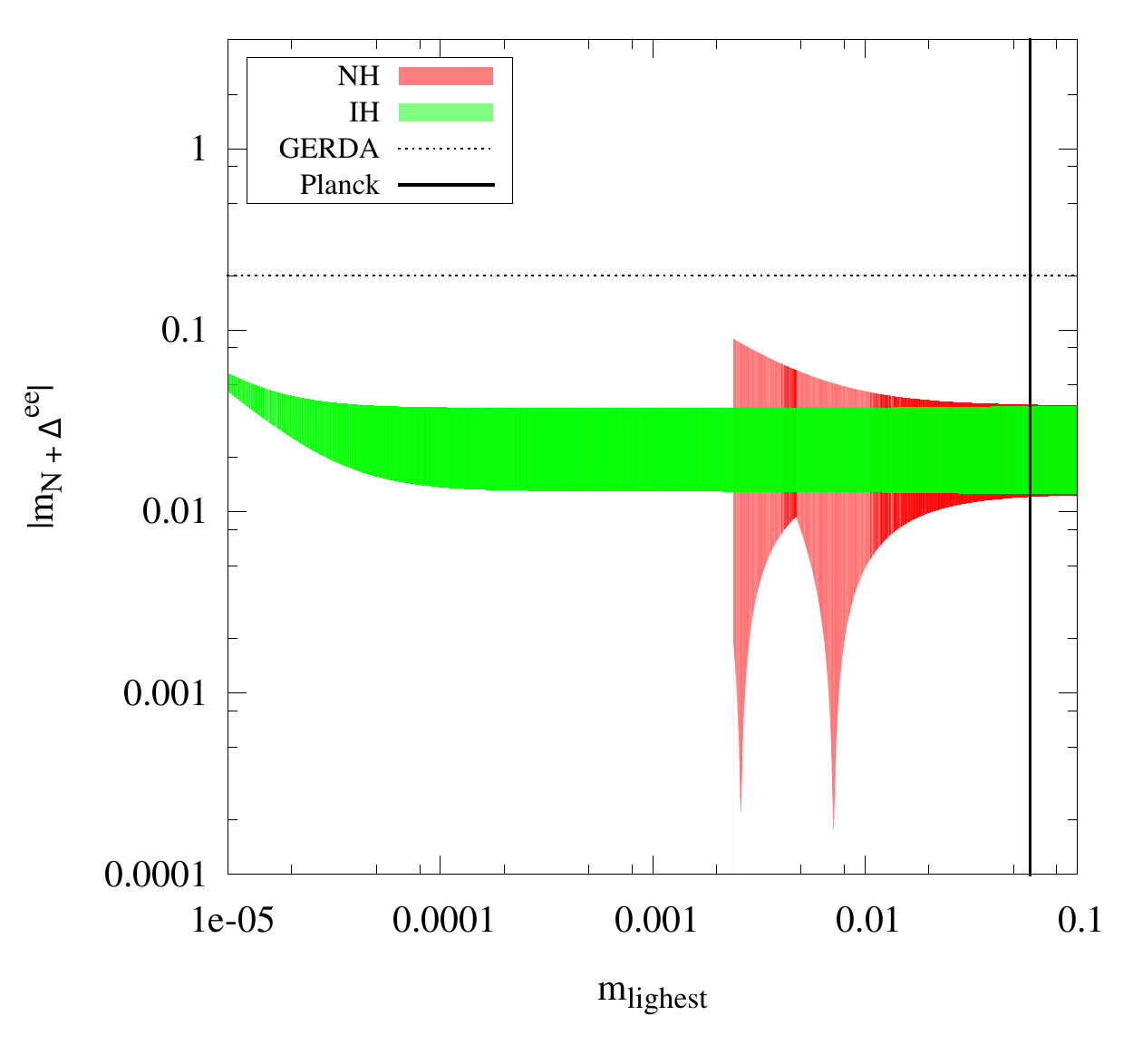,width=1.0\textwidth,clip=}
%\end{tabular}
\caption{New physics contribution to effective neutrino mass which appears in NDBD for the diagrams shown in figure \ref{fig0} with type I seesaw dominance and $M_{W_R}= 3.5$ TeV, $M_{\Delta_R} = 1$ TeV.}
\label{fig20}
\end{figure}
\begin{figure}[!h]
\centering
%\begin{tabular}{cc}
\epsfig{file=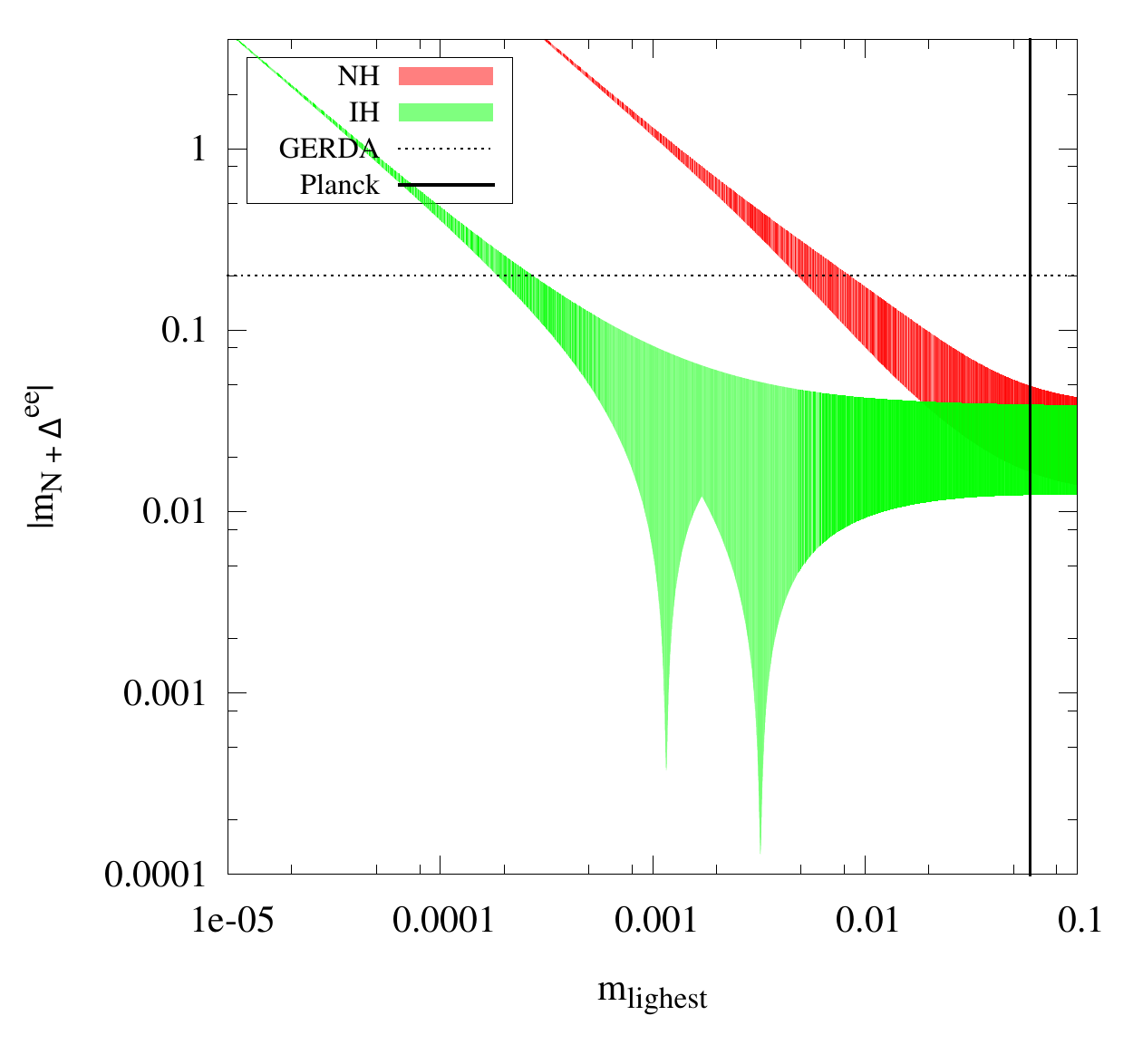,width=1.0\textwidth,clip=}
%\end{tabular}
\caption{New physics contribution to effective neutrino mass which appears in NDBD for the diagrams shown in figure \ref{fig0} with type II seesaw dominance and $M_{W_R}= 3.5$ TeV, $M_{\Delta_R} = 1$ TeV.}
\label{fig30}
\end{figure}

\section{Lepton Flavor Violation and Collider Bounds}
\label{sec:lfvlhc}
Lepton flavor violation (LFV) in LRSM were studied in details in previous works including \cite{LFVLR}. Within this model, there are several possible LFV processes like $\mu \rightarrow e\gamma, \mu \rightarrow 3e$. Here we consider $\mu \rightarrow 3e$ process mediated by doubly charged bosons in LRSM. It turns out that imposing the experimental bound on this process $\text{BR}(\mu \rightarrow 3e) < 10 \times 10^{-12}$ \cite{sindrum} is enough to keep other LFV processes within experimental limit. The branching ratio for the $\mu \rightarrow 3e$ process induced by doubly charged bosons $\Delta^{++}_L, \Delta^{++}_R$ is given by \cite{LFVLR}
\begin{equation}
\text{BR}(\mu \rightarrow 3e) = \frac{1}{2} \lvert h_{\mu e} h^{*}_{ee} \rvert^2 \left ( \frac{M^4_{W_L}}{M^4_{\Delta_L}}+\frac{M^4_{W_L}}{M^4_{\Delta_r}} \right )
\label{eqBR}
\end{equation}
where the couplings $h$ are given by
\begin{equation}
h_{ij} = \sum_n \left ( K_R \right )_{ni} \left ( K_R \right )_{nj} \sqrt{\left(\frac{M_n}{M_{W_R}}\right)^2}
\label{eqhij}
\end{equation}
In equation \eqref{eqBR}, $M_{\Delta_{L,R}}$ are the masses of $\Delta^{++}_{L,R}$ and in equation \eqref{eqhij}, $K_R$ is the right handed leptonic mixing matrix. In a previous work \cite{ndbd00}, the experimental bound on this LFV process was incorporated to restrict $M^{\text{heaviest}}_n/M_{\Delta}$, where $\frac{1}{M^2_{\Delta}} = \frac{1}{M^2_{\Delta_L}}+\frac{1}{M^2_{\Delta_R}}$. It was found that for most of the parameter space, $M^{\text{heaviest}}_n/M_{\Delta}<0.1$ with $M_{W_R} = 3.5$ TeV. Assuming $M_{\Delta_L} = M_{\Delta_R} = M_{\delta}$, the above bound will become $M^{\text{heaviest}}_n/M_{\delta}<0.1/\sqrt{2}$. However, this bound was calculated only with the assumption that $K_R = K_L$ and hence may not be applicable in a general case where both type I and type II seesaw terms contribute to light neutrino masses.

Apart from LFV bounds on the ratio $M^{\text{heaviest}}_n/M_{\Delta}$, there exists other experimental bounds on the new particles of LRSM. The most stringent bound on the additional charged vector boson $W_R$ comes from the $K-\bar{K}$ mixing: $M_{W_R} > 2.5$ TeV \cite{kkbar}. Direct searches at LHC also put similar constraints on the mass of $W_R$ boson. Dijet resonance search by ATLAS puts a bound $M_{W_R} > 2.45$ TeV at $95\%$ CL \cite{dijetATLAS}. This bound can however be relaxed to $M_{W_R} \geq 2$ TeV if $g_R \approx 0.6 g_L$. There are other bounds on $M_{W_R}$ coming from other searches in LHC experiments, but they are weaker than the dijet resonance bound and hence skipped here. Similarly, the doubly charged scalars also face limits from CMS and ATLAS experiments at LHC:
$$ M_{\Delta^{\pm \pm}} \geq 445 \; \text{GeV} \; (409 \; \text{GeV}) \; \text{for} \; \text{CMS (ATLAS)} $$
These limits have been put by assuming $100\%$ leptonic branching factions \cite{hdlhc}. The heavy right handed neutrinos with $SU(2)_R$ gauge interactions are also constrained by direct searches at LHC. For example, the search for $W_R \rightarrow l N$ at ATLAS and CMS constrains the right handed neutrino masses to be around 1 TeV \cite{rhnlhc}. All these experimental bounds are taken into account in our analysis below.
\section{Numerical Analysis}
\label{sec3}
We first calculate the standard light neutrino contribution to NDBD \eqref{eq:ndbd0} by evaluating the corresponding effective neutrino mass 
$$ m^{\text{eff}}_{\nu} = U^2_{Lei}m_i$$
where $U_L = U_{\text{PMNS}}$ is given by equation \eqref{matrixPMNS}. This is effectively the $(1,1)$ (or $(ee)$ in flavor basis) element of the light neutrino mass matrix $M_{\nu}=U_{\text{PMNS}}M^{\text{diag}}_{\nu} U^T_{\text{PMNS}}$ given as
\begin{equation}
m^{\text{eff}}_{\nu}=m^{\text{ee}}_{\nu}=m_1c^2_{12}c^2_{13}+m_2s^2_{12}c^2_{13}e^{2i\alpha}+m_3 s^2_{13}e^{2i\beta} 
\end{equation}
Using the best fit values of three mixing angle and two mass squared differences as shown in table \ref{param}, one can show the variation of $m^{\text{eff}}_{\nu}$ as a function of lightest neutrino mass $m_{\text{lightest}} = m_1 (\text{NH}), m_3 (\text{IH})$. This is shown in figure \ref{fig1} where the Majorana CP phases $\alpha, \beta$ are varied in the range $(0-2\pi)$, allowed by neutrino oscillation data. It is to be noted that the best fit neutrino parameters except $\delta$ in table \ref{param} are taken from \cite{valle14}. A value of Dirac CP phase $\delta = -\pi/2$ has been reported recently by experimental data \cite{diracphase}, but the standard light neutrino contribution to NDBD is independent of $\delta$ as seen from the above expression. It can be seen from figure \ref{fig1} that the light neutrino contribution can saturate the GERDA bound \cite{GERDA} only for higher values of lightest neutrino masses, disallowed by the Planck data on sum of absolute neutrino masses \cite{Planck13}. Thus, the Planck constraint on the parameter space shown in figure \ref{fig1} is more strict compared to the GERDA data as far as light neutrino contribution to NDBD is concerned. However, the GERDA limit is strong enough to rule out many new physics contributions as we discuss below.

For the new physics contribution discussed in this work, the total effective mass is 
\begin{align}
m^{\text{eff}}_{N+\Delta_R} &= \left[ p^2 \frac{M^2_{W_L}}{M^2_{W_R}}\frac{U^{*2}_{Rei}}{M_i} + p^2\frac{M^4_{W_L}}{M^4_{W_R}} \frac{U^2_{Rei}M_i}{M^2_{\Delta_R}}\right] \label{eq1} \\
\end{align}
which, for type I and type II seesaw dominance is calculated following the analysis discussed in the previous section. These are shown in figure \ref{fig2},\ref{fig3}, \ref{fig20} and \ref{fig30} respectively for two different values of $\Delta^{++}_{L,R}$ masses. For all these cases, the Dirac CP phase $\delta$ is varied in the allowed $3\sigma$ range $(0-2\pi)$. Also, $M_{W_R} = 3.5$ TeV and $M^{\text{heaviest}}_n/M_{\delta}=0.1/\sqrt{2}$. Since for either type I or type II dominating cases we are considering equality of left and right handed mixing matrices (as discussed above), the LFV bound $M^{\text{heaviest}}_n/M_{\delta}=0.1/\sqrt{2}$ is applicable here. It can be seen from figure \ref{fig2} that for type I seesaw dominance with $M_{\delta} = 3.5$ TeV, the new physics contribution to NDBD is sizable only when $m_{\text{lightest}} > 0.003$ eV for normal hierarchy. For inverted hierarchy however, the new physics contribution to NDBD is about ten times suppressed compared to the standard light neutrino contribution shown in figure \ref{fig1}. Similarly, for type II seesaw dominance with $M_{\delta} = 3.5$ TeV, GERDA upper limit rules out $m_{\text{lightest}}$ less than approximately $2 \times 10^{-3}$ eV and $5 \times 10^{-5}$ eV for NH and IH respectively, as seen in figure \ref{fig3}. Of course, these limits are for specific values of $M_{W_R}, M^{\text{heaviest}}_n/M_{\delta}$ mentioned above. We have also assumed $M_{\delta} = M_{W_R}$ for the cases shown in figure \ref{fig2}, \ref{fig3}. Considering a smaller value of $M_{\delta}$ say, 1 TeV and hence a smaller $M^{\text{heaviest}}$ lifts up the new physics contribution to NDBD as $m^{\text{eff}}_{N+\Delta_R}$ is inversely proportional to their masses, seen from equation \eqref{eq1}. This is shown in the $m^{\text{eff}}-m_{\text{lightest}}$ plane in figure \ref{fig20} and \ref{fig30}.

Now let us consider the interesting scenario where both type I and type II seesaw terms are equally dominating. In this case, the simple relation $(U_L=U_R)$ between the diagonalizing matrices of left and right handed sectors no longer holds, unlike the cases discussed above. Here we have to diagonalize the right handed neutrino mass matrix explicitly in order to calculate its eigenvalues $M_i$ as well as $U_R$. In the above equation \eqref{eq1},  $U_R$ is the diagonalizing matrix of $M_{RR}$ which, in terms of type II seesaw mass matrix can be written as 
\begin{equation}
M_{RR} = \frac{1}{\gamma}\left(\frac{v_R}{M_{W_L}}\right)^2 M^{II}_{\nu}
\label{eqmrr}
\end{equation}
Thus, the right handed neutrino masses are inversely proportional to the dimensionless parameter $\gamma$. Therefore, the two new physics contributions in the above equation \eqref{eq1} have different dependence on $\gamma$ as can be seen from the equation below.
\begin{align}
m^{\text{eff}}_{N+\Delta_R} &= \left[ \underbrace{p^2 \frac{M^2_{W_L}}{M^2_{W_R}}\frac{U^{*2}_{Rei}}{M_i}}_{\propto \gamma} + \underbrace{p^2\frac{M^4_{W_L}}{M^4_{W_R}} \frac{U^2_{Rei}M_i}{M^2_{\Delta_R}}}_{\propto \frac{1}{\gamma}}\right] 
\label{eq2gamma}
\end{align}
Although dimensionless parameter $\gamma$ should be of order one, in TeV scale type I+II seesaw, this has to be extremely fine-tuned. It is straightforward to see that for $M_i \approx 1$ TeV, $M_{W_R} \approx 3$ TeV, $m^{II}_{\nu} \approx 0.1 $ eV, the equation \eqref{eqmrr} gives $\gamma \approx 10^{-8}$. Keeping these parameters fixed, if we increase $\gamma$, then the type II seesaw contribution to light neutrino mass will exceed the desired range $\leq 0.1$ eV. This increase can however be compensated by increase in the strength (parameterized by $X$) of type I seesaw contribution so that difference between two large contributions from type I and type II seesaw can still give sub-eV scale light neutrino masses. 
\begin{figure}[!h]
\centering
\begin{tabular}{cc}
\epsfig{file=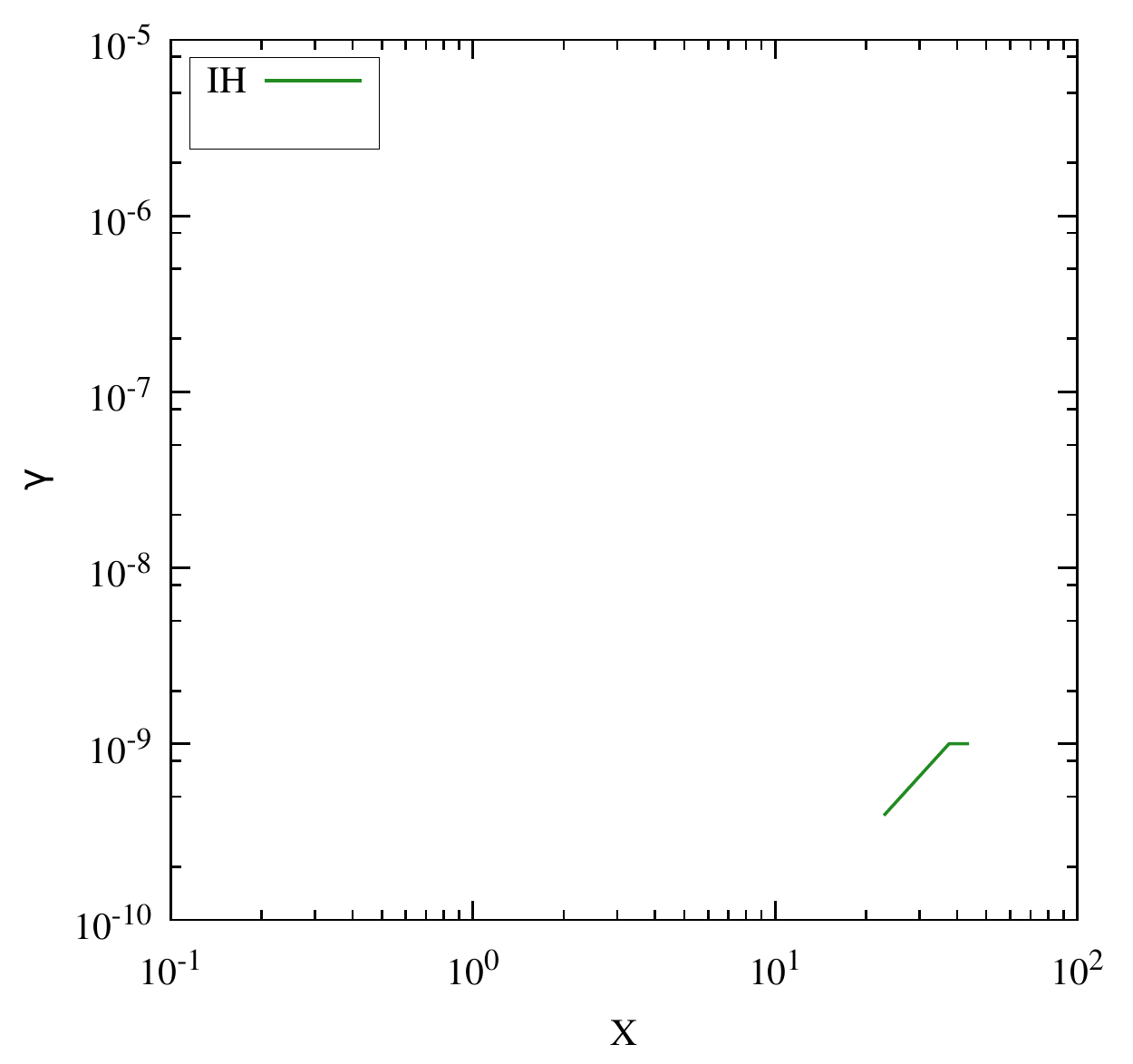,width=0.5\textwidth,clip=}&
\epsfig{file=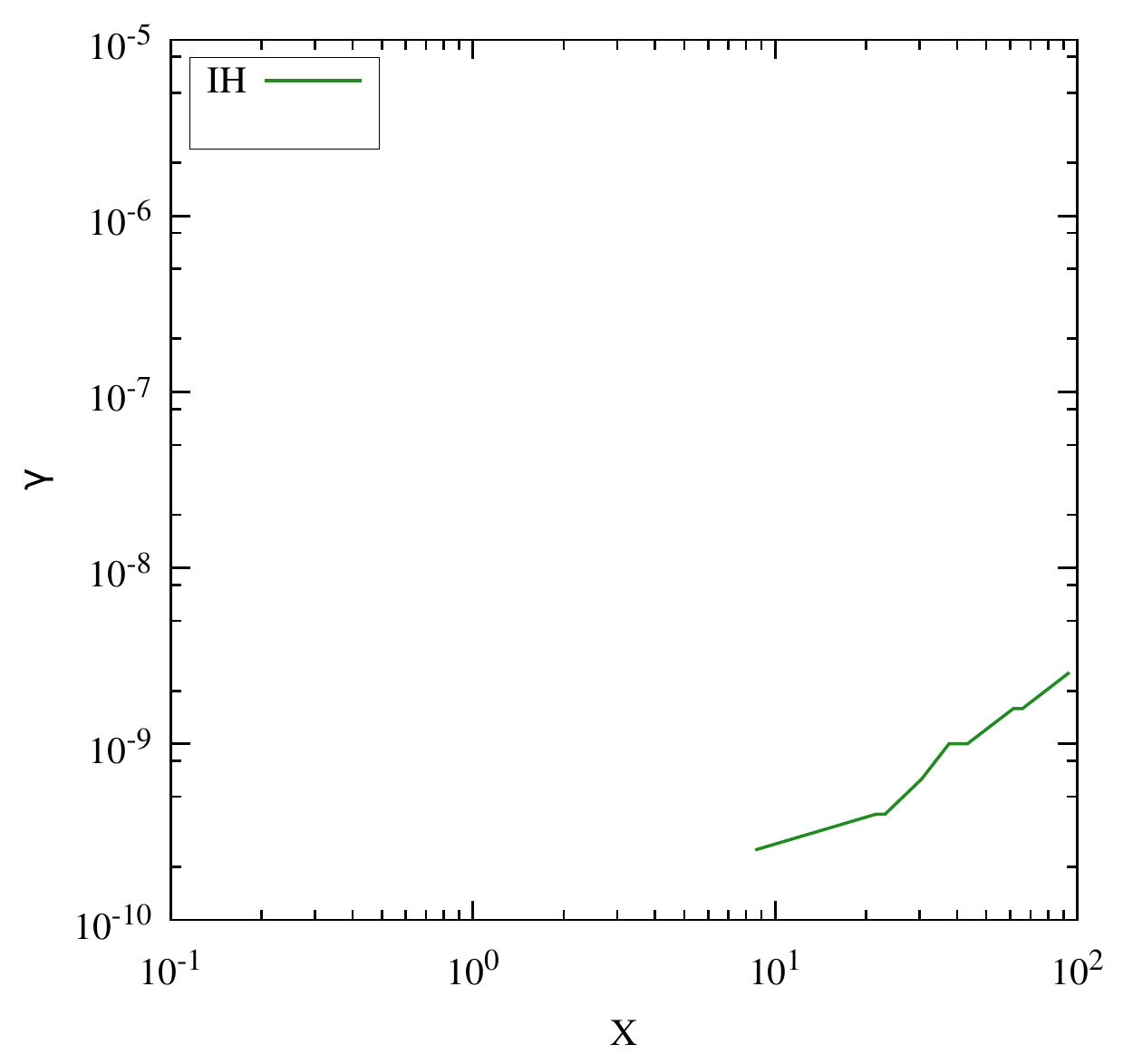,width=0.5\textwidth,clip=} \\
\end{tabular}
\caption{Constraints on dimensionless parameters $X$ and $\gamma$ from GERDA, Planck, LFV and perturbative bound for $M_{W_R} = 3.5$ TeV and $M_{\delta} = v_R, M^{\text{max}}_{\delta}$ respectively.}
\label{fig04}
\end{figure}
\begin{figure}[!h]
\centering
\begin{tabular}{cc}
\epsfig{file=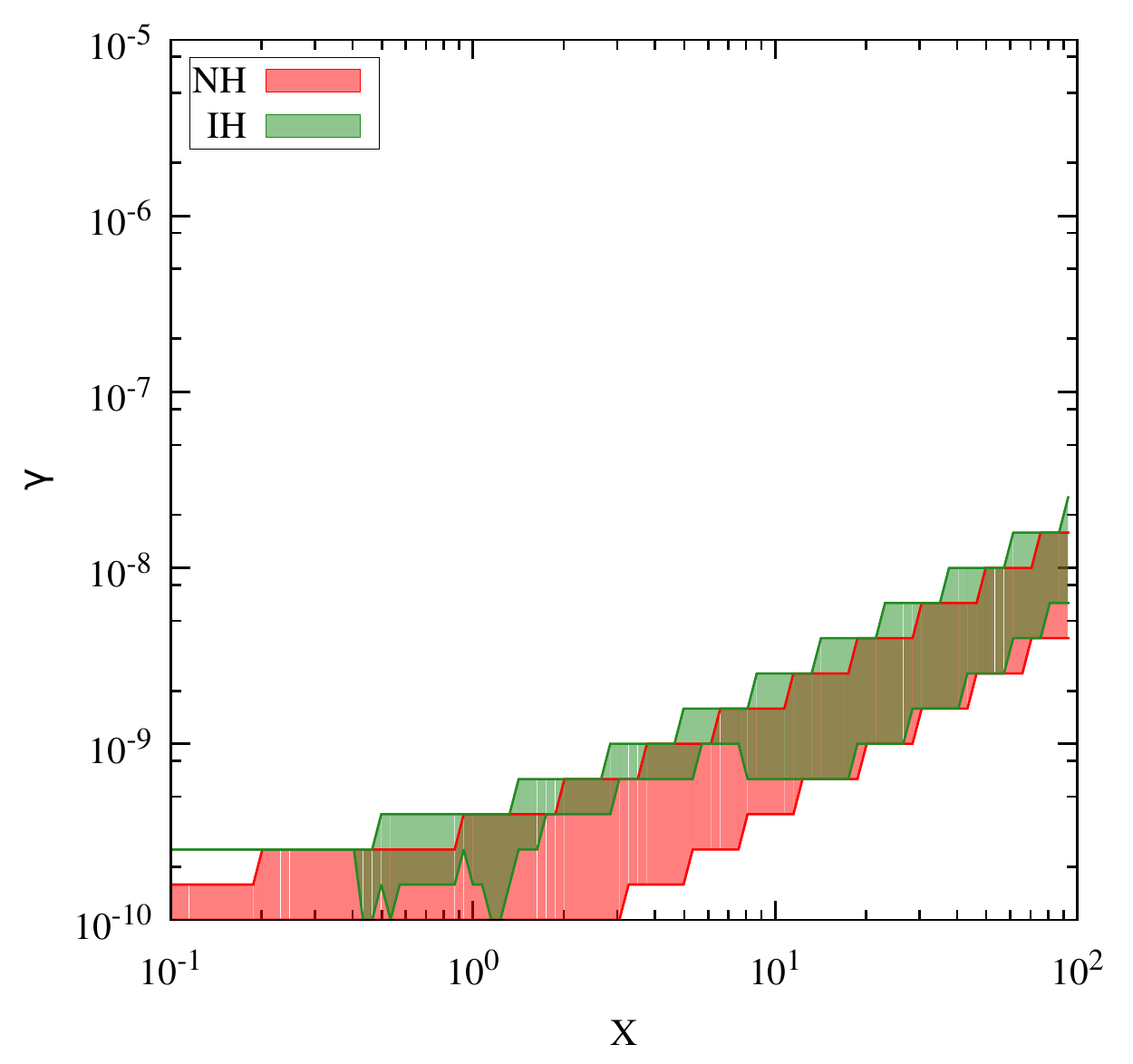,width=0.5\textwidth,clip=}&
\epsfig{file=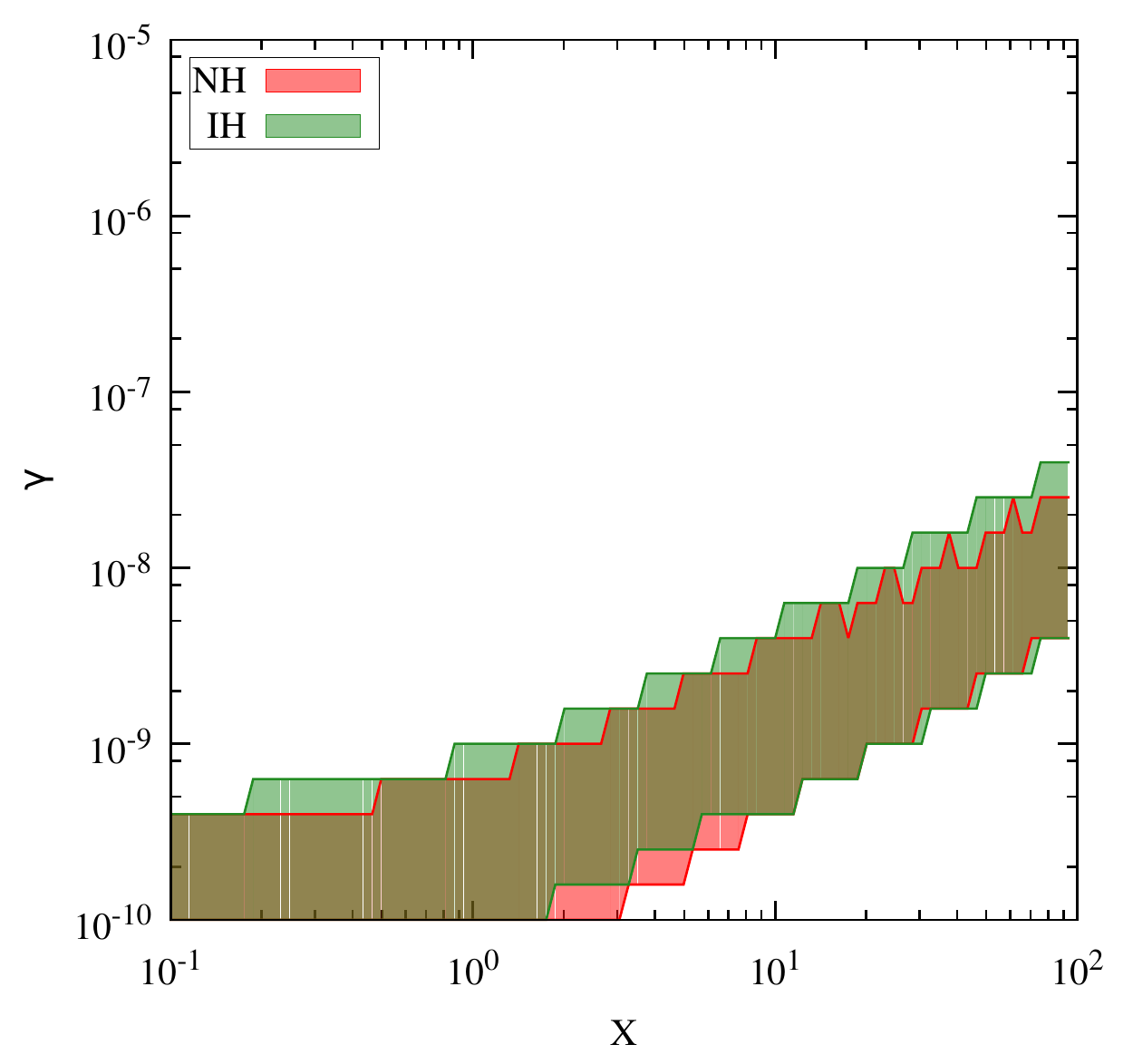,width=0.5\textwidth,clip=} \\
\end{tabular}
\caption{Constraints on dimensionless parameters $X$ and $\gamma$ from GERDA, Planck, LFV and perturbative bound for $M_{W_R} = 7$ TeV and $M_{\delta} = v_R, M^{\text{max}}_{\delta}$ respectively.}
\label{fig041}
\end{figure}
\begin{figure}[!h]
\centering
\begin{tabular}{cc}
\epsfig{file=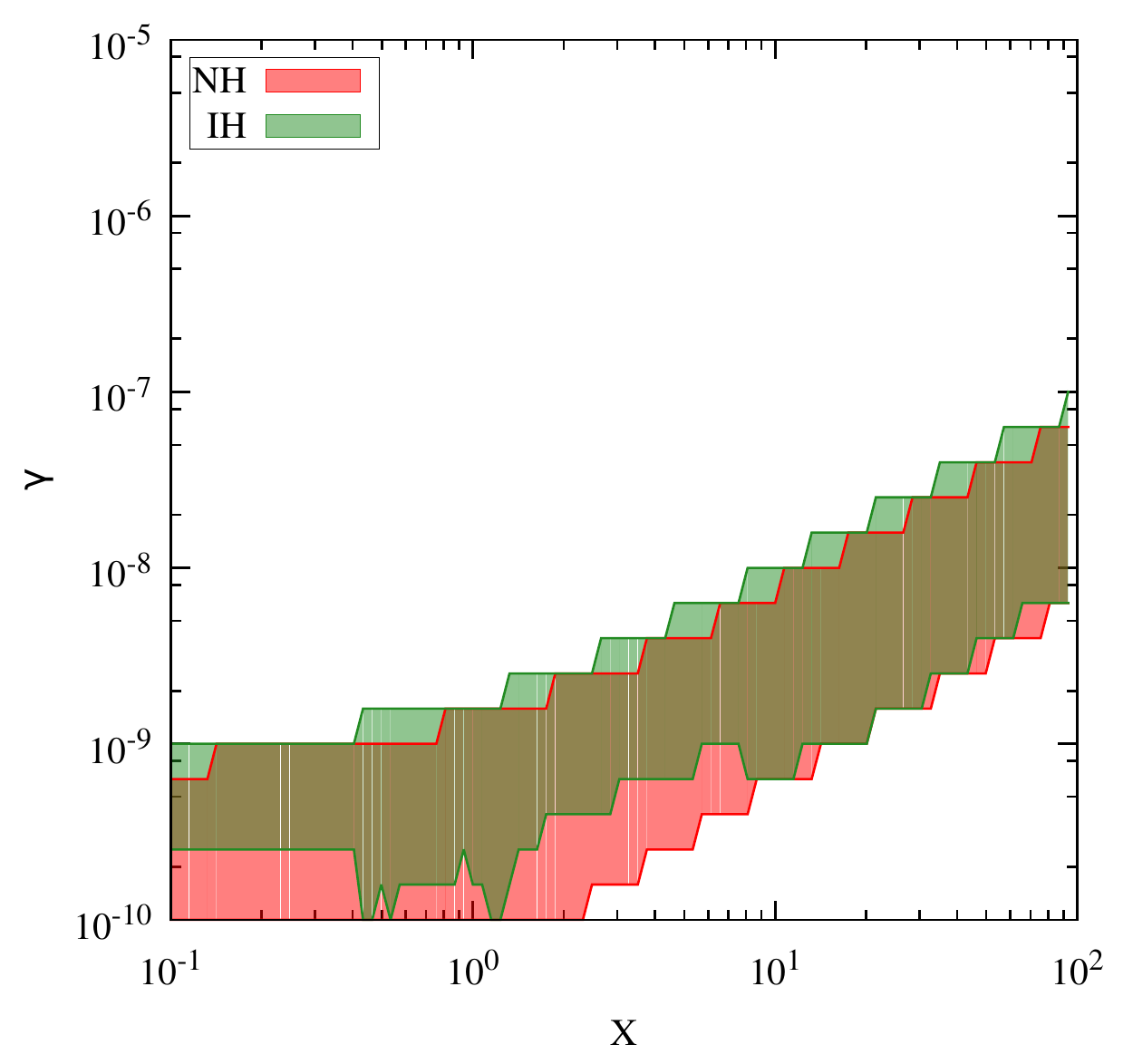,width=0.5\textwidth,clip=}&
\epsfig{file=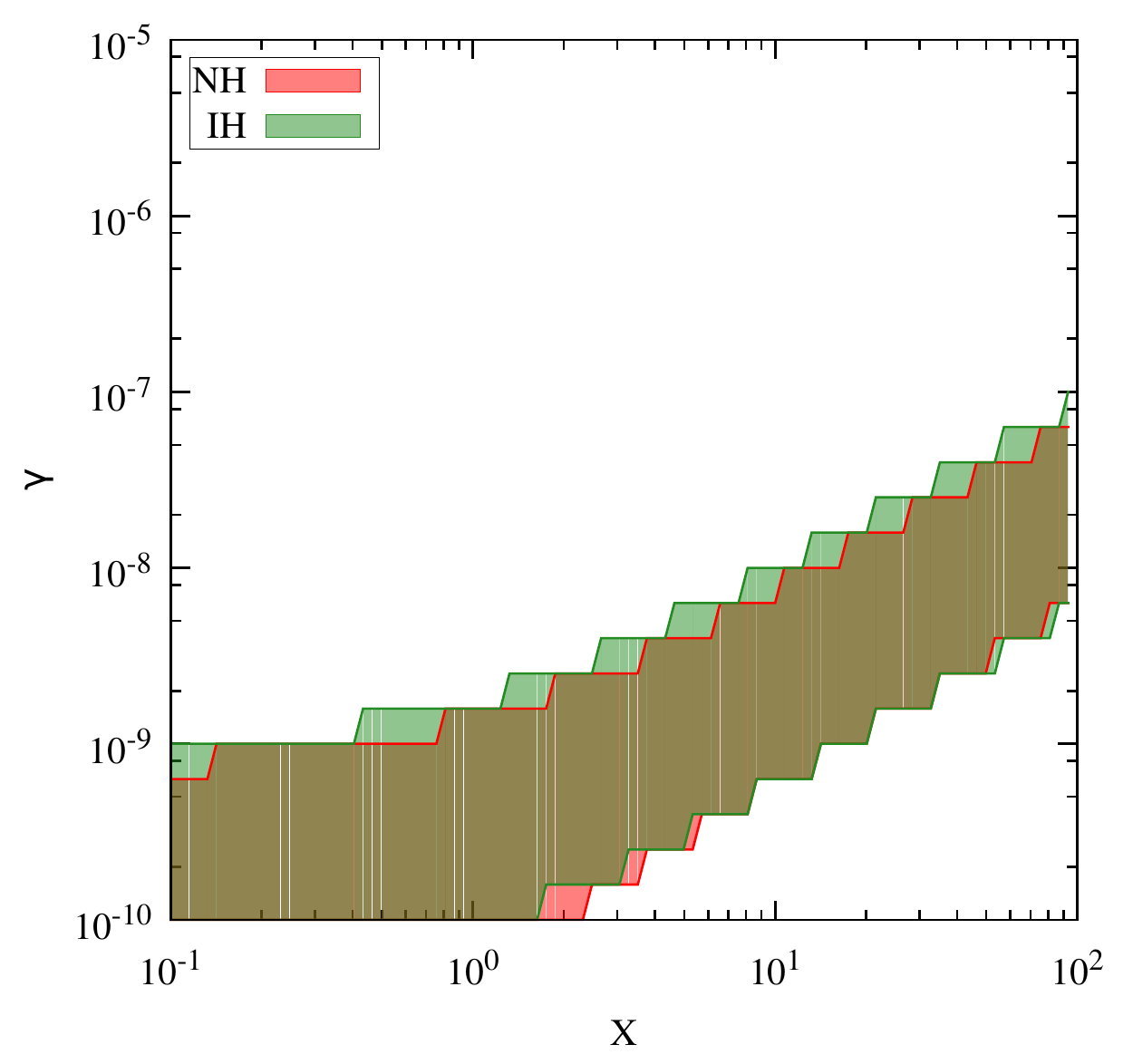,width=0.5\textwidth,clip=} \\
\end{tabular}
\caption{Constraints on dimensionless parameters $X$ and $\gamma$ from GERDA, Planck, LFV and perturbative bound for $M_{W_R} = 10$ TeV and $M_{\delta} = v_R, M^{\text{max}}_{\delta}$ respectively.}
\label{fig05}
\end{figure}
\begin{figure}[!h]
\centering
%\begin{tabular}{cc}
\epsfig{file=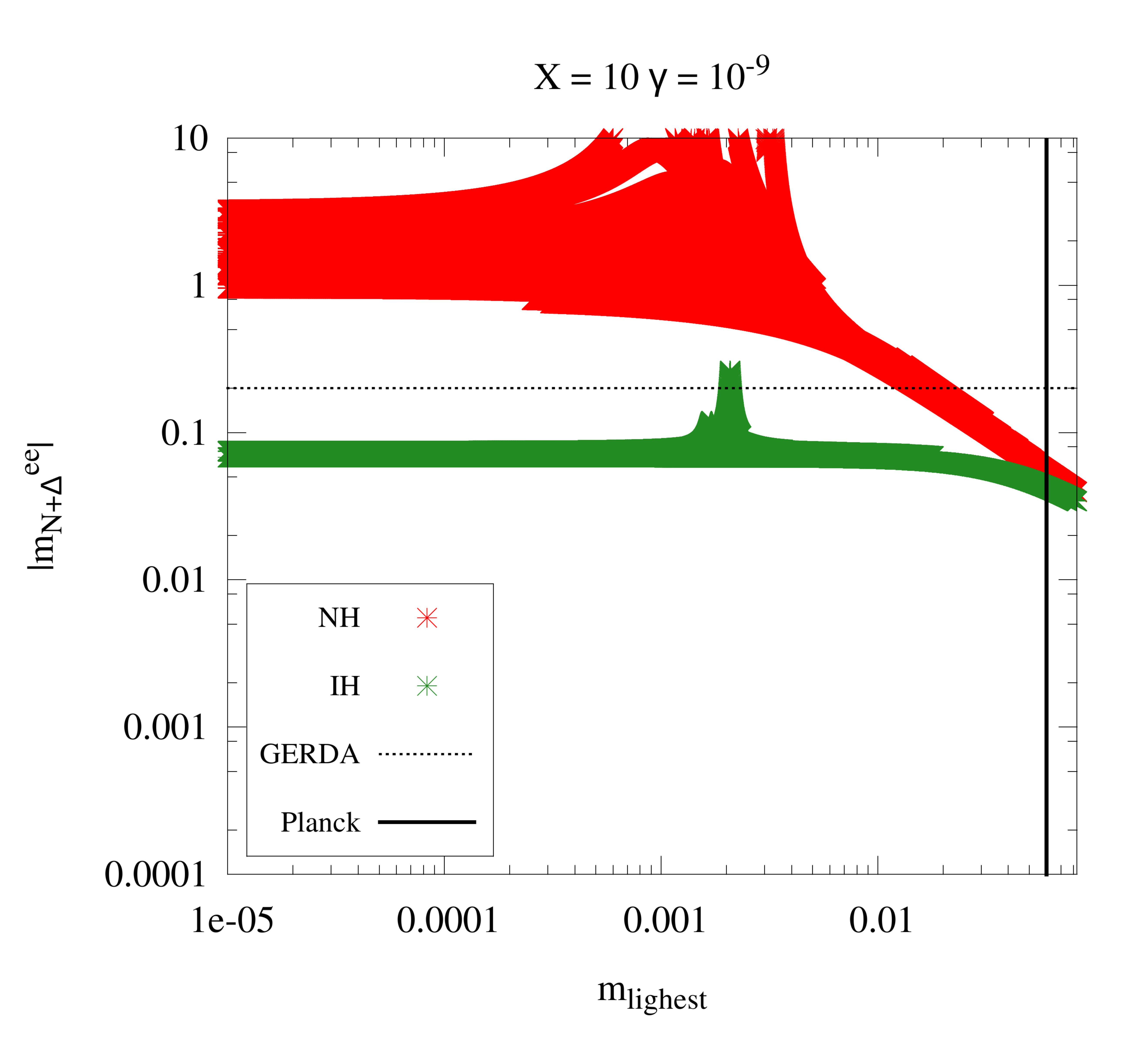,width=1.0\textwidth,clip=}
%\epsfig{file=X-1Y10.pdf,width=0.5\textwidth,clip=} \\
%\epsfig{file=X-1Y11.pdf,width=0.5\textwidth,clip=}&
%\epsfig{file=X-1Y12.pdf,width=0.5\textwidth,clip=} \\
%\end{tabular}
\caption{New physics contribution to effective neutrino mass which appears in NDBD for the diagrams shown in figure \ref{fig0} with type I+II seesaw for $M_{W_R} = 7$ TeV and $M_{\delta} = M^{\text{max}}_{\delta}$.}
\label{fig4}
\end{figure}
\begin{figure}[!h]
\centering
%\begin{tabular}{cc}
\epsfig{file=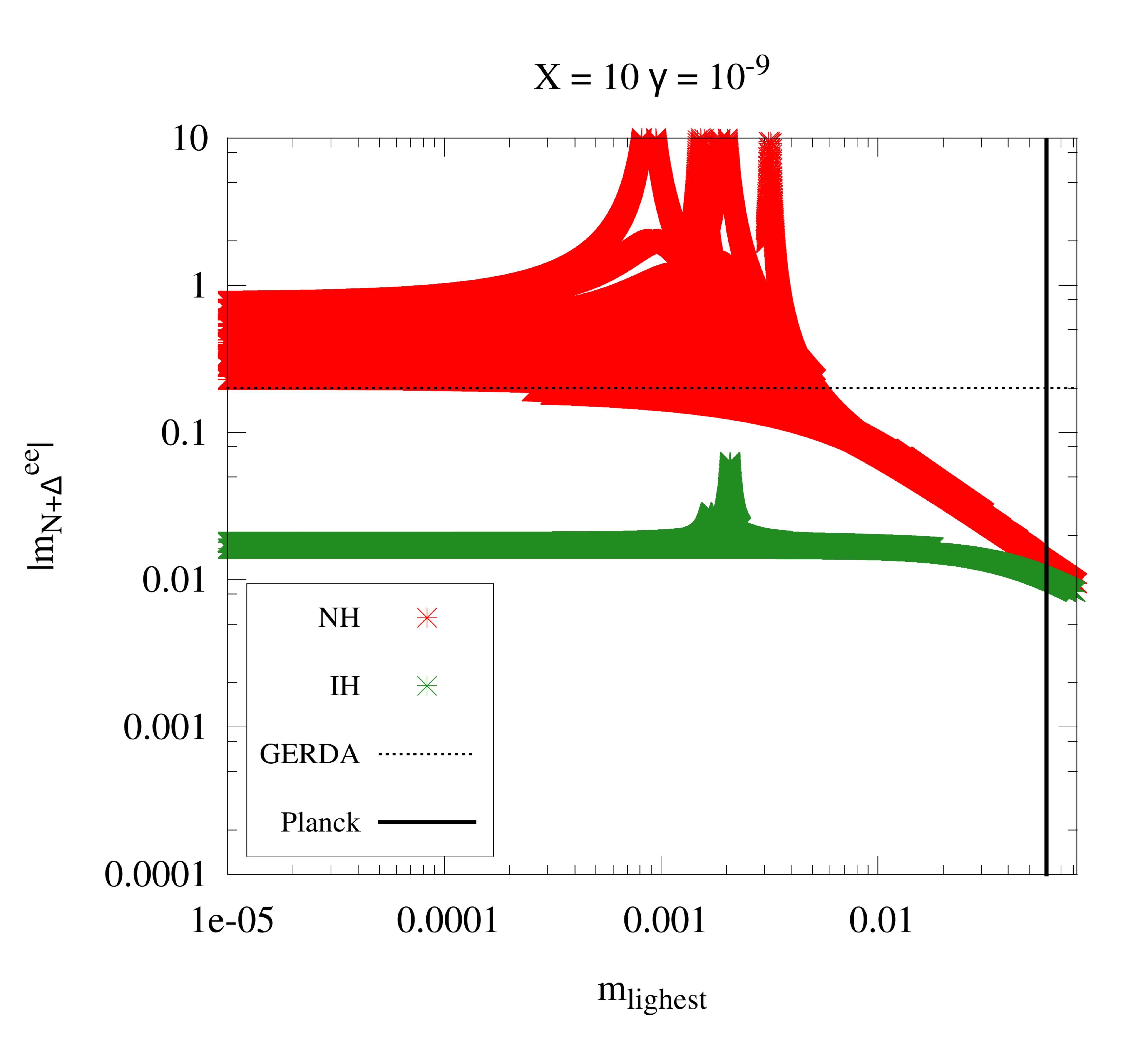,width=1.0\textwidth,clip=}
%\epsfig{file=X0Y10.pdf,width=0.5\textwidth,clip=} \\
%\epsfig{file=X0Y11.pdf,width=0.5\textwidth,clip=}&
%\epsfig{file=X0Y12.pdf,width=0.5\textwidth,clip=} \\
% \end{tabular}
 \caption{New physics contribution to effective neutrino mass which appears in NDBD for the diagrams shown in figure \ref{fig0} with type I+II seesaw for $M_{W_R} = 10$ TeV and $M_{\delta} = M^{\text{max}}_{\delta}$.}
 \label{fig5}
\end{figure}

%\begin{figure}[!h]
%\centering
%\begin{tabular}{cc}
%\epsfig{file=X2Y9.pdf,width=0.5\textwidth,clip=}&
%\epsfig{file=X2Y10.pdf,width=0.5\textwidth,clip=} \\
%\epsfig{file=X2Y11.pdf,width=0.5\textwidth,clip=}&
%\epsfig{file=X2Y12.pdf,width=0.5\textwidth,clip=} \\
%\end{tabular}
%\caption{Total contribution to effective neutrino mass which appears in NDBD for the diagrams shown in figure \ref{fig0} with type I+II seesaw}
%\label{fig7}
%\end{figure}

\begin{figure}[!h]
\centering
%\begin{tabular}{cc}
\epsfig{file=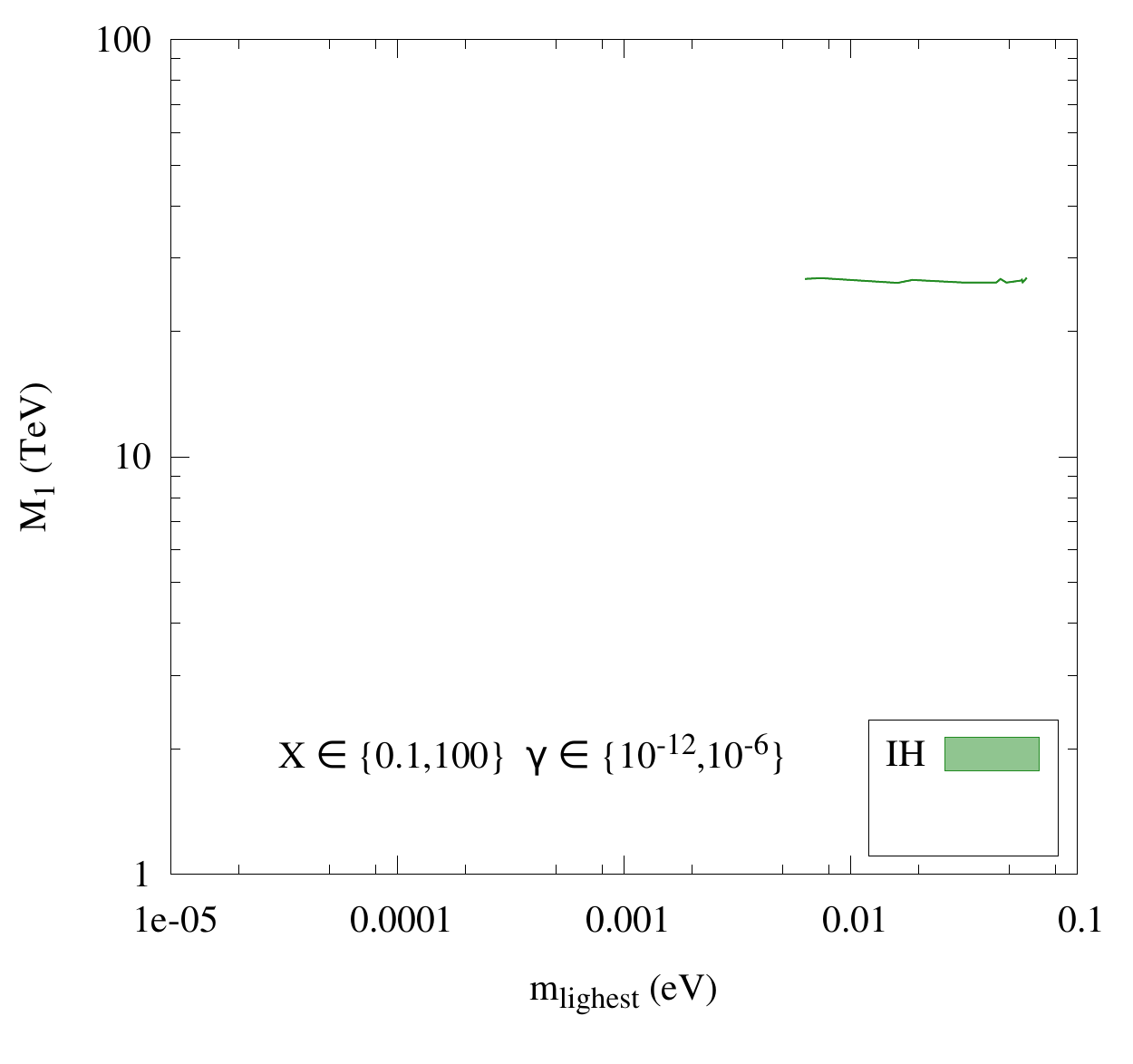,width=1.0\textwidth,clip=}
%\epsfig{file=mMI.pdf,width=0.5\textwidth,clip=} \\
%\end{tabular}
\caption{Constraints on the right handed neutrino mass in type I+II seesaw scenario for $M_{W_R} = 3.5$ TeV and $M_{\delta} = M^{\text{max}}_{\delta}$}
\label{fig9}
\end{figure}
\begin{figure}[!h]
\centering
%\begin{tabular}{cc}
\epsfig{file=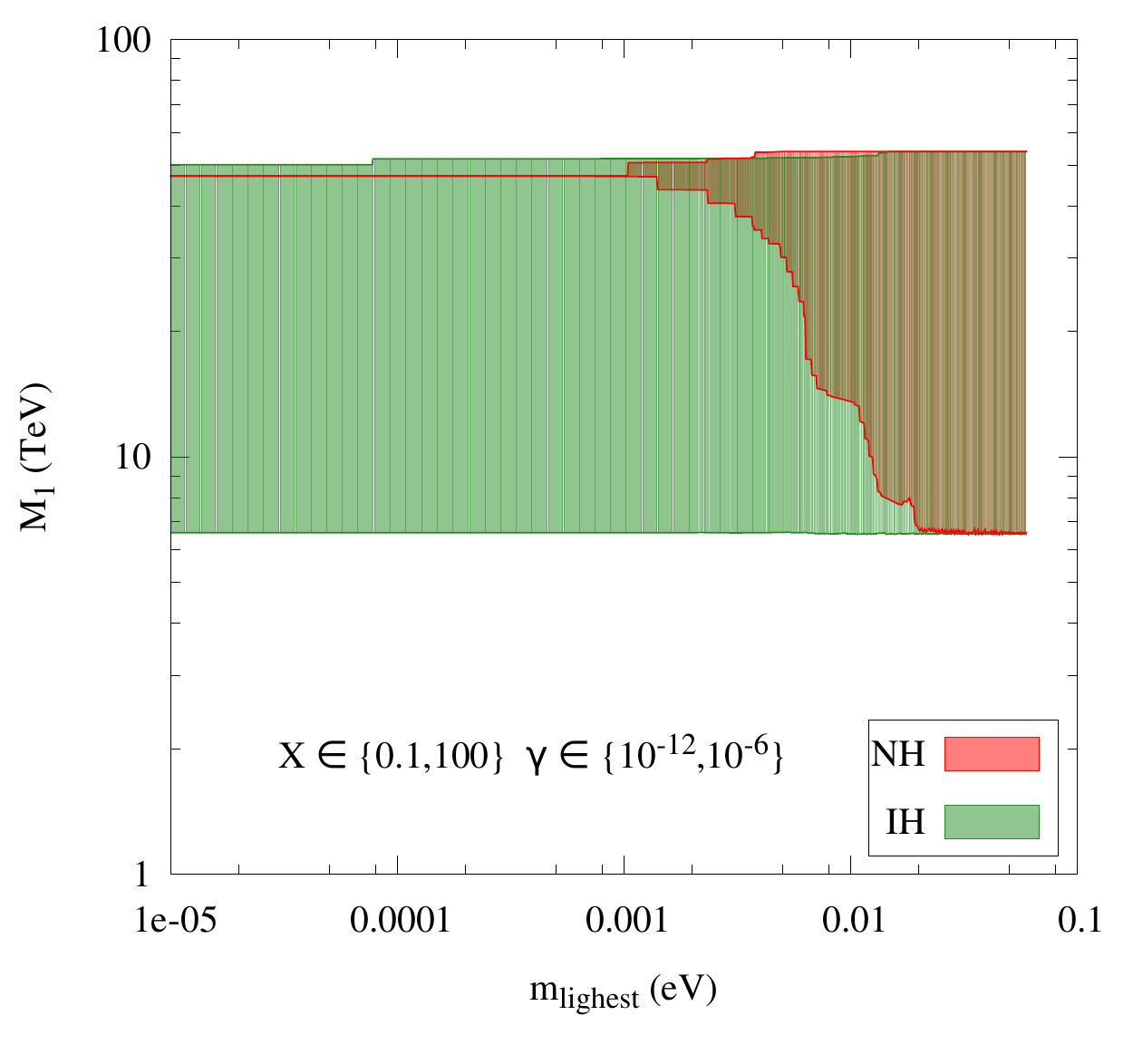,width=1.0\textwidth,clip=}
%\epsfig{file=mMI.pdf,width=0.5\textwidth,clip=} \\
%\end{tabular}
\caption{Constraints on the right handed neutrino mass in type I+II seesaw scenario for $M_{W_R} = 7$ TeV and $M_{\delta} = M^{\text{max}}_{\delta}$}
\label{fig10}
\end{figure}
\begin{figure}[!h]
\centering
%\begin{tabular}{cc}
\epsfig{file=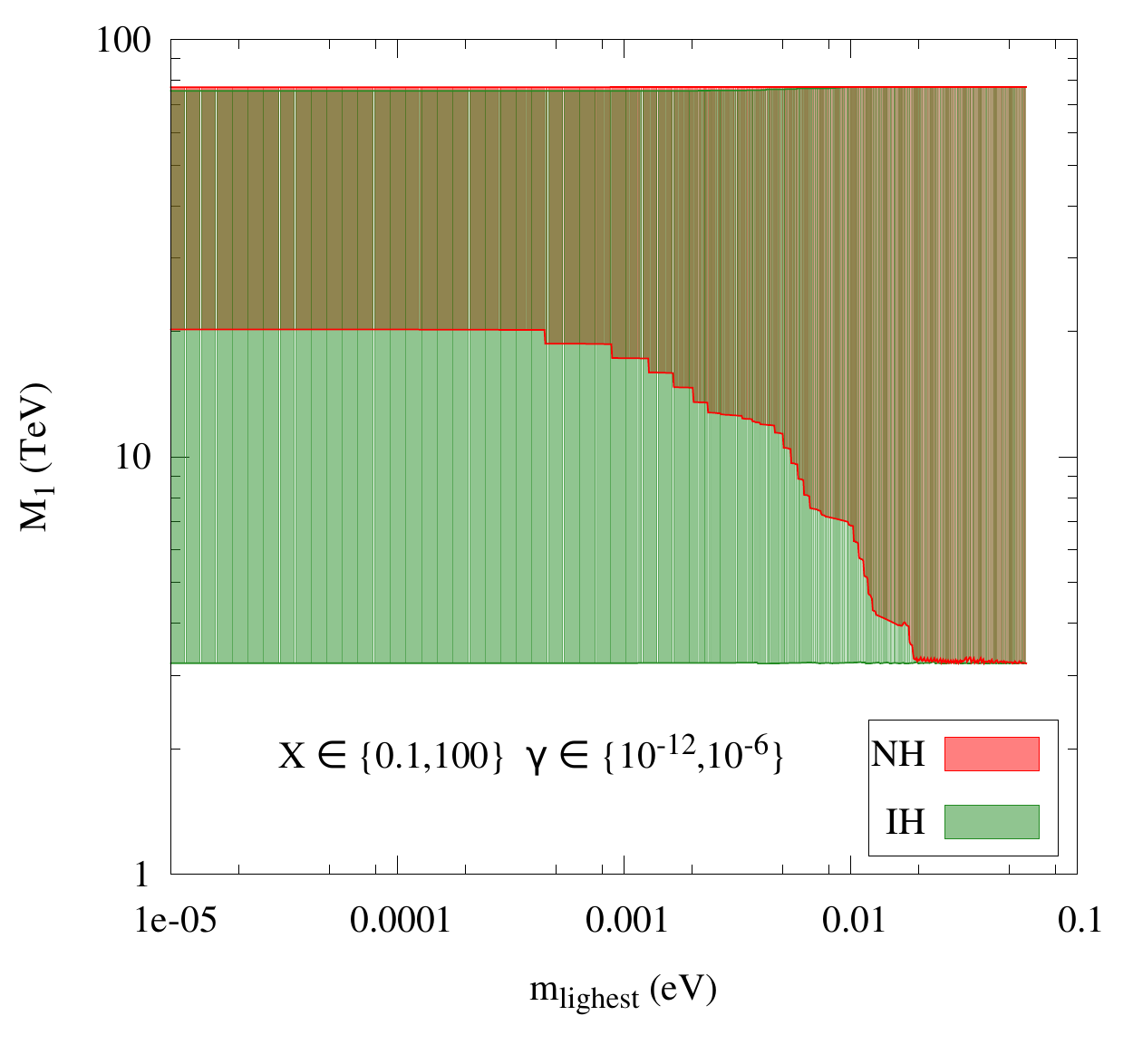,width=1.0\textwidth,clip=}
%\epsfig{file=mMI.pdf,width=0.5\textwidth,clip=} \\
%\end{tabular}
\caption{Constraints on the right handed neutrino mass in type I+II seesaw scenario for $M_{W_R} = 10$ TeV and $M_{\delta} = M^{\text{max}}_{\delta}$}
\label{fig11}
\end{figure}

Since the two new physics contributions have opposite dependences on $\gamma$ as seen from the equation \eqref{eq2gamma} above, it is interesting to see the effects of its variation on $m^{\text{eff}}$. We first vary the parameters $(\gamma, X)$continuously and calculate the effective neutrino mass for the specific parameter values given in table \ref{param}. For each chosen value of $M_{W_R}$, we consider three possible values of $M_{\delta}$ namely, $M_{W_R}, v_R$ and $M^{\text{max}}_{\delta}$. It is to be noted that the physical mass of the doubly charged component of $\Delta_R$ is given by $M^2_{\delta^{++}_{R}} \approx 2 \rho_2 v^2_R$ where $\rho_2$ is a dimensionless coupling in the scalar potential. Thus, the maximum mass squared of this doubly charged boson is $2 \rho_2^{\text{max}} v^2_R = (M^{\text{max}}_{\delta^{++}_{R}})^2$ where $\rho_{\text{max}} = \sqrt{4\pi}$ is the maximum value of dimensionless coupling allowed by perturbative unitarity bound. Also, the $W_R$ boson mass is given by $M^2_{W_R} \approx \frac{g^2_R v^2_R}{2}$. Thus, given a specific choice of $M_{W_R}$, the value of $v_R$ can be found out from this formula. We also assume equality of gauge couplings $g_L = g_R$ in accordance with the left-right symmetry.

We impose the GERDA bound on the new physics contribution to $m^{\text{eff}}$, Planck bound on the sum of absolute neutrino masses and LFV bound on $\text{BR}(\mu \rightarrow 3e)$. We explicitly calculate $\text{BR}(\mu \rightarrow 3e)$ using the expression given in equation \eqref{eqBR} without any assumptions about the right handed lepton mixing matrix and hence about the couplings $h_{ij}$ given in equation \eqref{eqhij}. We compute these couplings explicitly for a given right handed neutrino mass matrix given by equation \eqref{eqmrr}. Since we are calculating the right handed neutrino masses for a specific combination of $(\gamma, X)$, we also impose the constraint that the heaviest right handed neutrino mass is $f^{\text{max}} v_R$ where $f^{\text{max}} = \sqrt{4\pi}$, the perturbative upper bound on dimensionless couplings. After imposing these four bounds, we finally show the allowed parameter space in $(\gamma, X)$ plane. Figure \ref{fig04} shows the allowed parameter space for $M_{W_R} = 3.5$ TeV with two different values of $M_{\delta} = v_R, M^{\text{max}}_{\delta}$. It is clearly seen that only a few points with inverted hierarchy are allowed by the constraints imposed. Thus the contribution to NDBD will be dominated mostly by the standard light neutrino contribution in this case. We see that more and more regions of parameter space are allowed if we consider larger values of $M_{W_R}$ namely, 7 TeV and 10 TeV. They are shown in figure \ref{fig041} and figure \ref{fig05} for two possible values of $M_{\delta} = v_R, M^{\text{max}}_{\delta}$ like before. It can be seen from these plots that higher values of $\gamma$ are allowed by the constraints only if the parameter $X$ is taken to be high. This is equivalent to increasing the strength of type I seesaw term in the neutrino mass formula \eqref{type2}.

After finding the allowed parameter space in terms of $(\gamma, X)$ for three different choices of $M_{W_R}$, we show the variation of $m^{\text{eff}}$ (for new physics contributions) with lightest neutrino mass for a particular combination of $(\gamma, X)$. For example, we calculate $m^{\text{eff}}$ for $\gamma = 10^{-9}, X = 10$ with $M_{W_R} = 7$ TeV and $M_{\delta} = M^{\text{max}}_{\delta}$. Its variation with $m_{\text{lightest}}$ is shown in figure \ref{fig4}. It can be seen from this figure that most of the parameter space for NH is ruled out for $m_{\text{lightest}} < 0.01$ eV whereas most of the parameter space with IH are allowed by GERDA upper bound on $m^{\text{eff}}$. Similarly, we also calculate $m^{\text{eff}}$ with same combination of $(\gamma, X)$ but for $M_{W_R} = 10$ TeV and $M_{\delta} = M^{\text{max}}_{\delta}$. This is shown in figure \ref{fig5}. As expected, the contribution to $m^{\text{eff}}$ is suppressed compared to $M_{W_R} = 7$ TeV case due to heavier particles mediating NDBD. We choose the other parameter values from table \ref{param}. The right handed neutrino masses are explicitly calculated by diagonalizing the right handed neutrino mass matrix $M_{RR}$. After choosing $(\gamma, X)$ as well as the parameters given in table \ref{param}, the right handed neutrino mass matrix $M_{RR}$ contains four free parameters $(m_{\text{lightest}}, \delta, \alpha, \beta)$. We vary these parameters continuously and calculate $m^{\text{eff}}$ whose variation with $m_{\text{lightest}}$ can be seen from figure \ref{fig4} and \ref{fig5}. 

Similarly, we can also constrain the right handed neutrino masses, as for each combination of $(\gamma, X)$, there exists a corresponding $M_{RR}$ given by equation \eqref{eqmrr}. The lightest right handed neutrino mass allowed by all these bounds in type I+II seesaw scenario discussed in this work for three different values of $W_R$ mass $M_{W_R} = 3.5, 7, 10$ TeV and $M_{\delta} = M^{\text{max}}_{\delta}$ are shown in figure \ref{fig9}, \ref{fig10} and \ref{fig11} respectively. It should be noted that, these bounds have been found only by constraining the new physics contribution to NDBD from available experimental as well as perturbative unitarity bounds. 

\section{Results and Discussion}
\label{sec4}
We have studied the consequences of a combination of type I and type II seesaw on the amplitude of neutrinoless double beta decay within the framework of a left-right symmetric model. Due to the presence of additional gauge bosons and scalars, this model has several new sources of neutrinoless double beta decay. We have considered these extra gauge bosons and scalars to be near a few TeV to strengthen their contributions to the amplitude of NDBD. For simplicity, we have assumed zero mixing between right handed and left handed gauge bosons as well as between heavy and light neutrinos. This results in three important contributions to NDBD shown in figure \ref{fig0}. Within this simplified setup, we first show the standard light neutrino contribution to the amplitude of NDBD. After this, the new physics contributions to NDBD is calculated assuming either type I or type II seesaw to contribute fully to the light neutrino masses and compare with the standard light neutrino contribution to NDBD. These are shown in figure \ref{fig1}, \ref{fig2}, \ref{fig3}, \ref{fig20} and \ref{fig30} respectively for different choices of $M_{W_R}, M_{\delta}$. We have also incorporated the LFV bounds coming from the experimental search for $\mu \rightarrow e\gamma, \mu \rightarrow 3e$ in the calculation.

After confirming the results of several earlier works in this seesaw limit, we then move onto the interesting case where both type I and type II seesaw terms can equally contribute to light neutrino masses. This allows us to choose the type I and type II seesaw mass matrices arbitrarily, provided they sum up to give the correct light neutrino mass matrix. Instead of considering such arbitrary mass matrices, we consider a very specific type I seesaw mass matrix which gives rise to tri-bimaximal type neutrino mixing with $\theta_{12} \simeq 35.3^o, \; \theta_{23} = 45^o$ and $\theta_{13} = 0$. The strength of the type I seesaw term is parametrized by a dimensionless parameter $X$. The type II seesaw mass matrix is them evaluated in terms of light neutrino mass matrix, constructed using the best fit neutrino data and the TBM form of type I seesaw mass matrix. The type II mass matrix therefore, is written in terms of five free parameters: the two unknown Majorana CP phases, one Dirac CP phase, the lightest neutrino mass and the dimensionless parameter $X$. In LRSM, the type II seesaw mass matrix is also proportional to the right handed neutrino mass matrix, which allows us to construct $M_{RR}$ in terms of the already derived type II seesaw mass matrix upto a constant of proportionality. After choosing the other related parameters as given in table \ref{param}, the constant of proportionality between type II seesaw mass matrix and $M_{RR}$ is $\gamma$ as seen from equation \eqref{eqmrr}. We show from equation \eqref{eqmrr} that for $M^{II}_{\nu}$ of the order of light neutrino masses, and lightest right handed neutrino mass around a TeV, the dimensionless parameter $\gamma$ has to be fine tuned around $10^{-8}$. We scan the parameter space $\gamma, X$ from the requirement of keeping new physics contribution to $m^{\text{ee}}$ below the GERDA upper limit as well as to satisfy the LFV and perturbative bounds. The allowed regions are shown in figure \ref{fig04}, \ref{fig041} and \ref{fig05} for three different values of $W_R$ masses. One can see that larger values of $\gamma$ are allowed only when $X$ is also increased. This behavior can be understood by looking at the formula for $m^{\text{ee}}$ given by equation \eqref{eq2gamma}. The first term within brackets on the right hand side of \eqref{eq2gamma} is inversely proportional to right handed neutrino masses. From equation \eqref{eqmrr}, the right handed neutrino mass is inversely proportional to $\gamma$ and directly proportional to type II seesaw term. Therefore, the $N-W^-_R$ contribution to $m^{\text{ee}}$ is directly proportional to $\gamma$ and inversely proportional to type II seesaw term. From the expressions of type II seesaw mass matrix given in appendix \ref{appendix1}, it can be seen that $M^{II}_{\nu}$ increases with increasing $X$. Therefore, $m^{\text{eff}}_{N}$ is directly proportional to $\gamma$ and inversely proportional to $X$ and hence increase in $\gamma$ has to be compensated by an increase in $X$ so as to keep $m^{\text{eff}}_{N}$ below the GERDA upper limit. Similar analysis can also be made for the second term within brackets on the right hand side of equation \eqref{eq2gamma}.

We then consider a pair of benchmark values of the parameters $\gamma = 10^{-9}, X=10$ and calculate the new physics contribution to the effective neutrino mass $m^{\text{eff}} = m^{\text{ee}}$ of NDBD. The variations of $m^{\text{ee}}$ with lightest neutrino mass are shown in figure \ref{fig4} and \ref{fig5} for $M_{W_R} = 7, 10$ TeV respectively. This also allows one to discriminate between light neutrino mass hierarchies. Since a particular choice of $(\gamma, X)$ fixes the right handed neutrino mass matrix upto the Majorana CP phases, one Dirac CP phase and $m_{\text{lightest}}$, one can also find out the allowed values of right handed neutrino masses for the allowed values of $(\gamma, X)$. We have shown the allowed values of lightest right handed neutrino mass $M_1$ with $m_{\text{lightest}}$ in figure \ref{fig9}, \ref{fig10} and \ref{fig11}. We note that with respect to the new physics contribution to NDBD, the allowed region of parameter space for $M_{W_R} = 3.5$ TeV is very small, implying that the standard light neutrino exchange contribution dominates. As $M_{W_R}$ is increased, the LFV and perturbative bounds become weaker and allows more region of parameter space in terms of $\gamma, X$ which can give rise to sizable new physics contributions to NDBD. It should be noted that the purely type I or purely type II seesaw cases discussed earlier had sizable new physics contributions to NDBD. However, they can not be considered as limiting cases of the general type I+II seesaw discussed here. This is due to the fact that, $\gamma = 0$ will give type I seesaw dominance but with a TBM type mixing matrix, ruled out by experimental data.

We also note that the allowed values of the dimensionless parameter $\gamma$ given by equation \eqref{eq:gammaLR} are very tiny $(\sim 10^{-8}$ if only TeV scale type II seesaw dominates. However, in a framework with both type I and type II seesaw, $\gamma$ can be larger if the strength of type I seesaw $X$is also increased. Increasing $\gamma$ also reduces the right handed neutrino masses as clear from equation \eqref{eqmrr} which is also equivalent to increasing type I seesaw term, which is inversely proportional to right handed neutrino masses. Such tiny values of $\gamma$ are unnatural in most LRSM with TeV scale type II seesaw mechanism, and demands the role of some new physics behind it. Due to necessity of such unnatural fine-tuning of $\gamma$, there have been many studies of LRSM where the terms in the scalar potential leading to the $\beta$ terms appearing in the expression for $\gamma$ in equation \eqref{eq:gammaLR} are removed by imposing some symmetries, leading to $v_L = 0$ from the minimization of the scalar potential. This will give rise to a purely type I seesaw framework with TeV scale left-right symmetry. Such possibilities were discussed in the last reference of \cite{lrsm} and also in \cite{gammaLR}.

We have considered a very simplified picture of neutrinoless double beta decay in left-right symmetric model ignoring the contributions from left-right mixing as well as heavy-light neutrino mixing. However, we have pointed out the differences in $m^{\text{eff}}$ for individual seesaw dominance and equal dominance of both type I and type II seesaw, and constrained the model parameters in a way not considered before. However, a more detailed analysis taking into account all the new physics contributions to NDBD in LRSM should be pursued to give a more general conclusion. We also did not construct the UV complete flavor symmetry framework giving rise to the TBM form of type I seesaw mass matrix. It is undoubtedly a non-trivial exercise to implement discrete flavor symmetries in left-right symmetric models due to the difference in gauge structure and lepton representations from that in the standard model. Very recently such a model building work appeared in \cite{wernerA4}. We leave such a detailed analysis to a subsequent work.

\begin{acknowledgments}
The authors would like to thank Werner Rodejohann, MPIK, Heidelberg for pointing out a mistake in the first arXiv version of this work.
\end{acknowledgments}
%\clearpage

\appendix
\section{Elements of Type II Seesaw Mass Matrix}
\label{appendix1}

\begin{equation}
T_{11}=\left(c^2_{12}c^2_{13}+\frac{2X}{3}\right)m_1-\frac{1}{3}e^{2i\alpha}\left(-X-3s^2_{12}c^2_{13}\right)m_2+s^2_{13}e^{2i\beta}m_3
\end{equation}
\begin{align}
T_{12}=\frac{1}{3}((-X-3c_{12}c_{13}s_{12}c_{23}-3e^{i\delta}s_{13}s_{23}c^2_{12}c_{13})m_1-e^{2i\alpha}(-X-3s_{12}c_{12}c_{13}c_{23} \nonumber \\ 
+3e^{i\delta}s^2_{12}s_{13}s_{23}c_{13})m_2) 
+\frac{1}{3}(3s_{13}s_{23}c_{13}e^{i(2\beta+\delta)}m_3)
\end{align}
\begin{align}
T_{13}=\frac{1}{3}((-X+3s_{12}s_{23}c_{12}c_{13}-3e^{i\delta}s_{13}c^2_{12}c_{13}c_{23})m_1-e^{2i\alpha}(-X+3s_{12}s_{23}c_{12}c_{13} \nonumber \\ 
+3e^{i\delta}s^2_{12}s_{13}c_{13}c_{23})m_2)+\frac{1}{3}(3s_{13}c_{13}c_{23}e^{i(2\beta+\delta)}m_3)
\end{align}
\begin{align}
 T_{22}=\left( \left(s_{12}c_{23}+e^{i\delta}s_{13}s_{23}c_{12}\right)^{2}+\frac{X}{6}\right)m_1+\left(\left(c_{12}c_{23}-e^{i\delta}s_{12}s_{13}s_{23}\right)^2+\frac{X}{3}\right)e^{2i\alpha}m_2 \nonumber\\ +\left(s^2_{23}c^2_{13}+\frac{X}{2}\right)e^{2i(\beta +\delta)}m_3
\end{align}
\begin{align}
 T_{23}=\left( \left(c_{23}s_{12}+e^{i\delta}s_{13}s_{23}c_{12}\right)\left(-s_{12}s_{23}+e^{i\delta}s_{13}c_{12}c_{23}\right)+\frac{X}{6}\right)m_1\nonumber\\-\left(\left(c_{12}s_{23}+e^{i\delta}s_{12}s_{13}c_{23}\right)\left(c_{12}c_{23}-e^{i\delta}s_{12}s_{13}s_{23}\right)-\frac{X}{3}\right)e^{2i\alpha}m_2 \nonumber\\ +\left(s_{23}c^2_{13}c_{23}-\frac{X}{2}\right)e^{2i(\beta +\delta)}m_3
\end{align}
\begin{align}
T_{33}=\left( \left(-s_{12}s_{23}+e^{i\delta}s_{13}c_{12}c_{23}\right)^{2}+\frac{X}{6}\right)m_1+\left(\left(c_{12}s_{23}+e^{i\delta}s_{12}s_{13}c_{23}\right)^2+\frac{X}{3}\right)e^{2i\alpha}m_2\nonumber \\ +\left(c^2_{13}c^2_{23}+\frac{X}{2}\right)e^{2i(\beta +\delta)}m_3
\end{align}

\end{document}